\documentclass[prb, aps, twocolumn, asmath, amssymb, floatfix, superscriptaddress,longbibliography]{revtex4-2}
\usepackage[dvips]{graphics}
\usepackage[justification=raggedright]{caption}
\usepackage{subcaption}
\usepackage[font=footnotesize,figurewithin=none]{caption}
\usepackage{adjustbox}
\usepackage[labelfont=bf]{caption}
\usepackage[labelformat=parens, labelfont=bf]{subcaption}
\usepackage{color}
\usepackage{float}
\usepackage{natbib}
\usepackage[dvipsnames]{xcolor}
\setcitestyle{numbers}
\setcitestyle{square}
\definecolor{dred}{rgb}{0,0,0.6}
\definecolor{NavyBlue}{rgb}{0, 0, 128}
\usepackage{graphicx}
\usepackage{bm}
\usepackage{amssymb}
\usepackage{amsmath}
\usepackage{mathtools}
\usepackage{dcolumn}   
\usepackage[colorlinks, linkcolor=OrangeRed,citecolor=RoyalBlue,urlcolor=NavyBlue]{hyperref}
\usepackage[all]{hypcap} 
\usepackage{physics}
\usepackage{enumerate}
\usepackage{braket}        
\usepackage{overpic}
\usepackage{amsfonts}
\usepackage{dsfont}
\usepackage[mathscr]{eucal}
\usepackage{hyperref}
\usepackage{setspace}

\newcommand{\dg}{\dagger}
\newcommand{\mb}{\mathbf}

\newcommand{\prbsubref}[2]{\hyperref[#1]{\ref*{#1}{(\subref*{#2})}}}

\newcommand{\tc}[1]{\textcolor{OrangeRed}{#1}}
\begin{document}
\title{Magnons on a dice lattice: topological features and transport properties}

\author{Shreya Debnath and Saurabh Basu\\ \textit{Department of Physics, Indian Institute of Technology Guwahati, Guwahati-781039, Assam, India}}

\begin{abstract}
In this paper, we study the topological properties of magnons on a dice lattice, also known as the dual of a more widely studied kagome lattice. This structure has a central atom at the center of the honeycomb lattice, which leads to the formation of a flat band. Magnetic Hamiltonians associated with the magnon bands are scarcely studied in this flat band system, which motivates us on examining an interplay of different magnetic spin interactions, such as the Heisenberg exchange, Dzyaloshinskii-Moriya interaction (DMI), pseudodipolar interaction (PDI) and magnetocrystalline anisotropies in a dice lattice. In particular, the objective is to ascertain their roles in inducing various topological phases and the phase transitions therein. The magnons are obtained via a Holstein-Primakeoff transformation for a ferromagnetic Heisenberg model on a dice lattice, that renders an analogue of a pseudospin-1 fermionic Hamiltonian and yields higher Chern numbers, along with larger values of the thermal Hall conductivity. The competing effects of the DMI and the PDI in inducing transitions from either topological to topological or topological to trivial phases are noted and the corresponding results are supported via the magnon band structures, presence (or absence) of edge modes in a nanoribbon geometry, and the transport characteristic, namely the discontinuities in the thermal Hall conductivities. Meanwhile, the magnetocrystalline anisotropy also plays an intriguing role, where distinct (nonuniform) values at the different sublattice sites result in a richer topological landscape with Chern numbers $C = \pm 2$, $\pm 1$, and $0$, while a uniform anisotropy yields only $C = \pm 2$ and $0$. This discrepancy arises from the broken valley symmetry in the nonuniform case. Finally, we have enriched our understanding on the role of the \textit{flat} band in impacting the topological features by comparing some of the key results with that of a honeycomb structure.

\end{abstract}

\maketitle


\section{INTRODUCTION}
In this era of spintronic devices, studying nontrivial topological properties in magnetic materials has become a growing area of interest. Numerous studies have explored nontrivial phases in spin liquids~\cite{PhysRevX.14.021053}, skyrmions~\cite{skyrmion3_2022,skyrmion4_2021,skyrmion5_2023}, and ordered magnetic materials~\cite{moulsdale2019unconventional, zhuo2021topological, li2021magnonic, czajka2023planar}, along with spintronic devices~\cite{spintronics1_2018, spintronics2_2022,spintronics4_2023,spintronics5_2020,spintronics6_2019}. Moreover, magnetic materials provide an ideal platform to exhibit various transport phenomena, such as thermal Hall effect~\cite{thermalHallantiferro, sugii2017thermal, hirokane2019phononic, li2020phonon, thermalhall2010}, spin Seebeck effect~\cite{Seebeck}, magnon spin Nernst effect~\cite{NernstEffect1, NernstEffect2, NernstEffect3}, etc. In particular, there has been a plethora of work on magnons which are the collective excitations of magnetic systems. They behave like bosons and are charge neutral. Besides magnons, there exist various kinds of bosonic excitations that host topological phases, such as acoustic modes (phonon)~\cite{PhononHall}, electromagnetic waves (photon)~\cite{PhotonHall}, etc. Thus, apart from being a proprietary property of the electronic system, much attention has also been paid to these charge neutral excitations, which have subsequently opened the gateway to spin-based technologies~\cite{magnon-skyrmion,magnon1,skyrmion1,skyrmion2}. In this work, among other things, we shall benchmark the topological phases of the magnetic system to some of the analogous phenomena in the fermionic counterpart.

In magnetic systems, various types of magnetic interactions are involved, such as Heisenberg exchange interaction, magnetic anisotropy, Zeeman coupling, etc. Moreover, in presence of strong spin-orbit coupling, Dzyaloshinskii-Moriya interaction (DMI) and the pseudodipolar interaction (PDI) play crutial roles~\cite{PhysRevLett.69.836, PhysRevLett.102.017205}, which not only define the magnetic properties but also contribute the key elements in observing nontrivial band topology.  While DMI is dominant in a system that lacks inversion symmetry and favours chiral spin configurations, PDI is strong, where the strength of the spin-orbit coupling is much larger than the exchange interactions between the two isospins~\cite{PhysRevLett.102.017205}. The effects of either DMI or the PDI in topological magnon insulators have been studied earlier in various types of lattice models, such as square~\cite{debnath2023multiple, square1}, honeycomb~\cite{honeycomb1}, kagome~\cite{PhysRevB.98.094419, PhysRevX.10.041059, PhysRevB.89.134409}, Pyrochlore~\cite{PhysRevLett.118.177201}  lattices etc. However, in this work, we wish to explore the combined effects of DMI and PDI in inducing various topological phases and the phase transitions therein. We also demonstrate the interplay of these two interaction potentials on the transport properties, such as the thermal Hall effect, etc. The results are expected to be more interesting in the context of multiband systems, specifically if there is a flat band that can promote the study of strong interaction effects.

\begin{figure}[t]
    \centering
    \includegraphics[width=0.95\columnwidth]{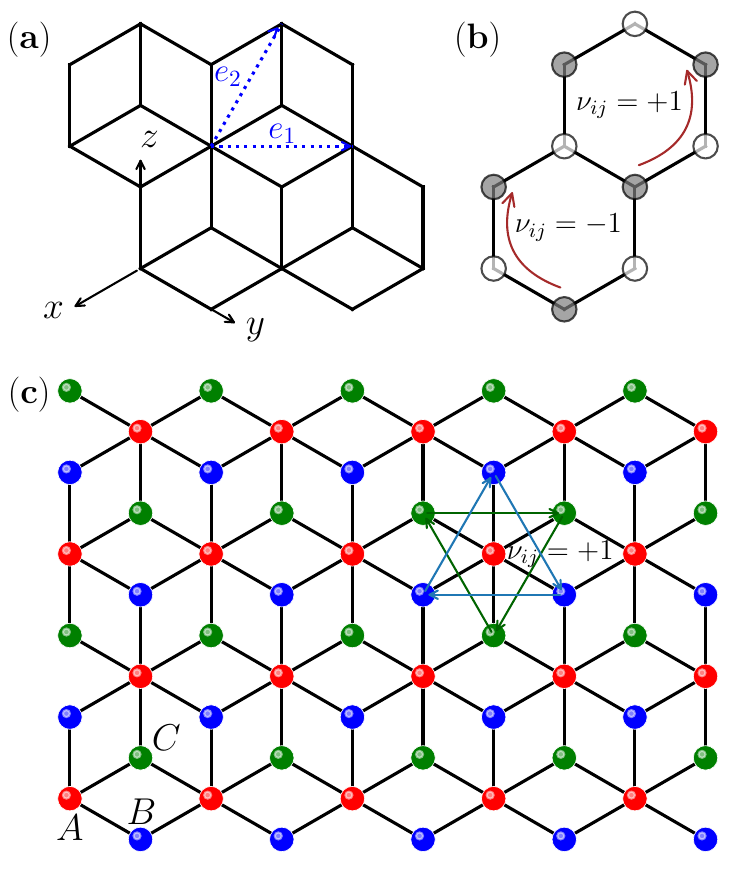}
    \caption{Schematic diagram for dice lattice, (a) where three adjacent (111) layers have been shown along with the directions $x$, $y$, $z$. $\mb{e}_1$ and $\mb{e}_2$ are two translation vectors. (b) A honeycomb structure has been shown to visualise the direction of the DMI with $\nu_{ij}$. $\nu_{ij}$ will be $+1$($-1$) for anticlockwise(clockwise) ordering of interactions between two identical sublattices via the other one.  (C) Schematic view of a dice lattice where red, blue, and green dots denote A, B, and C sublattices respectively. The effective DMI acts between the B and the C sublattices.}
    \label{Dice}
\end{figure}
In this study, we shall consider a dice lattice which is a bipartite lattice with an unequal number of two subsets of sites. While the coordination number of one subset (containing two sites) is 3, the other (containing one site) is twice of it, that is 6~\cite{PhysRevB.84.241103}, as shown in Fig.~\ref{Dice}\tc{(a)}. In other words, we consider a model consisting of three sublattices: A, B, and C. In this model, the A sublattice is connected to both the B and C sublattices with a coordination number of 6, while the B and C sublattices are not connected to each other. The system possesses inversion symmetry with respect to the A sublattice. Thus, a honeycomb structure with a central atom, mimicking a \textit{dice} has a possible experimental realization via appropriate sectioning of a cubic lattice and thus enables us for the exploration of its magnetic properties and topology. Such a crystal structure may be formed via three adjacent (111) layers of a cubic lattice. Hence, if a heterostructure is grown along (111) direction, the cubic symmetry is reduced in a trigonal symmetry, where $\mb{e}_1$ and $\mb{e}_2$ are the two translation vectors, as shown in Fig.~\ref{Dice}\tc{(a)}. 

Studying quantum Hall effect in fermionic systems has been investigated widely on dice lattices. Owing to the presence of a larger number of bands, including a flat band~\cite{PhysRevB.107.035421}, they subsequently exhibit higher Chern numbers, which corresponds to a larger value of anomalous quantized Hall conductivity. There are physical realization of this lattice structure, such as TMO (111) heterostructure~\cite{xiao2011interface}, transition metal oxide $\text{SrTiO}_3/\text{SrIrO}_3/\text{SrTiO}_3$ trilayer heterostructure~\cite{PhysRevB.84.241103}, etc, which have an electronic dispersion similar to that of a dice structure comprising of a conduction, valance and flat bands and hence, can capture some of the features mentioned above. We have embedded an interacting spin model on a dice lattice, and for the sake of studying its properties, we have transformed it into a particle (bosonic) Hamiltonian. The resulting model has a structure similar to that of an electronic system, which permits investigation of the band properties and hence defines their topological attributes. Although the magnetic properties of a dice lattice have been studied as a two-dimensional quantum magnet~\cite{dice_magnet, dice_magnet1}, to the best of our knowledge the topological aspects have not yet been explored. With magnons being charge neutral, probing the traditional Hall effect in a magnon insulator does not have a meaning (remember the Hall plateaus are quantized in unit of $\frac{e^2}{h}$, $e$: electronic charge), while studying thermal Hall conductivity for our system makes perfect sense~\cite{PhysRevB.99.014427, kim2024thermal}. In particular, we study the anomalous version of it. The description, as we shall see later, still survives in the presence of an external Magnetic field. 

It is well known that, finite Hall response requires broken time-reversal symmetry (TRS). Unlike fermionic systems which require violation of TRS via external agencies, such as a magentic field or a complex second neighbour hopping on a honeycomb lattice (the so called Haldane term), in magnetic systems, the order parameter, that is, the magnetization reverses its direction if time reverses; thereby violating the TRS. It is of further interest, that if DMI acts between the next nearest neighbor (NNN) lattice sites, which will be defined in the following section, it behaves similarly to the Haldane term in breaking TRS in the effective Hamiltonian~\cite{mcclarty2022topological}. As we shall deal with ferromagnetic interactions in the following discussions, according to magnetic space group symmetry, we introduce an effective TRS operator~\cite{magnonboundpairs2023}, namely, $\tau' = R(n, \pi)\tau$, which is a combination of the time reversal operator $\tau$ plus a rotation operator $R(n, \pi)$. Here, $n$ denotes the axis in the spin space, perpendicular to the spin orientation, with $\pi$ representing the angle of rotation about this axis. Thus, it preserves the spin state despite suffering a flip due to the TRS operator. Although the DMI remains unaffected under the $\tau$ operator, the rotation operator affects the DM vectors (defined later) by changing its direction. Hence, its presence breaks the effective TRS. Moreover, there exists a $U(1)$ symmetry, which implies the spin conservation or the particle number conservation, that is, $[H, N]=0$, where $H$ is generic particle Hamiltonian and $N$ is the number operator given by, $N=\sum a_i^\dagger a_i$~\cite{magnonboundpairs2023}. $a_i^\dagger$($a_i$) are the creation(annihilation) operators corresponding to the $i^{\text{th}}$ site on the lattice. In presence of nearest neighbour (NN) PDI, the system violates $U(1)$ symmetry or the spin conservation, leading to a nontrivial band topology. Additionally, the PDI depends on the NN bond direction and we shall show later that such a scenario renders the flat band to be weakly dispersive, enabling it to contribute to the thermal Hall conductivity~\cite{PhysRevB.107.035421}.

Due to the presence of a higher coordination number of A sublattice (see Fig.~\ref{Dice}), the sublattice symmetry is already broken. However, the system is still inversion symmetric about the A sublattice sites. To break this, we apply a nonuniform magnetic anisotropy at all the A, B, and C sublattices, which is able to induce a band gap in the magnon spectrum without the presence of DMI or PDI. The analogy with the fermionic scenario can be established via a staggered mass term $M$ being added to $M\hat{S}_z$ in a pseudospin-1 fermionic system~\cite{PhysRevB.107.245150}. This breaks the valley degeneracy~\cite{PhysRevB.110.045426, PhysRevB.107.035421} and it holds the reputation of inducing topological phase transitions via coupling with an external magnetic field. Thus, while a differing chemical potential may generate a staggered mass in an electronic system~\cite{PhysRevB.107.245150}, in a magnetic system a similar effect arises due to the presence of a nonuniform magnetic or magnetic crystalline anisotropy at different sublattice sites. In this paper, we shall focus on how both the DMI and the PDI and such a magnetic anisotropy engage in an intricate interplay to influence the topological and transport properties of a magnonic insulator.

The paper is organised as follows. In Sec.~\ref{Formalism}, we construct the model Hamiltonian for dice lattice with spin-spin interaction. In Sec.~\ref{uniform}, we study the combined effects of DMI and PDI on the spectral properties and extract their ramifications on the topological aspects via computing the topological phase diagram based on the knowledge of the Chern number. In Sec.~\ref{edge_modes}, we investigate corresponding magnonic edge modes in a semi-periodic nanoribbon geometry. In Sec.~\ref{thermal}, we study the transport properties, that is the thermal Hall conductivity. Further, in Sec.~\ref{Honeycomb_Lattice} to provide a comprehensive understanding, we briefly discuss an analogous scenario for a honeycomb geometry (like graphene). We finally conclude in Sec.~\ref{conclusion}.

\section{MODEL AND FORMALISM}
\label{Formalism}
We consider a ferromagnetic spin Hamiltonian which can be mapped onto a bipartite dice lattice model and may be written as,
\begin{equation}
    \begin{split}
        &H = - J \sum_{\braket{i,j}} \left(\mb{S}_i \cdot \mb{S}_j\right) - F \sum_{\braket{i,j}}  \left(\mb{S}_i\cdot \mb{e}_{ij}\right)\left(\mb{S}_j\cdot \mb{e}_{ij}\right) \\
        & + D\sum_{\braket{\braket{i,j}}}\nu_{ij}\hat{z}\left(\mb{S}_i \times \mb{S}_j\right)  -\frac{1}{2}\sum_i K_iS_{zi}^2 - \mu_B B_0 \sum_i S_{zi}.
    \end{split}\label{Hamiltonian}
\end{equation}
The different parameters are defined as follows.
$J$ is the NN ferromagnetic interaction ($J>0$), and $F$ is the NN pseudodipolar interaction strength that arises due to the presence of strong spin-orbit coupling. Additionally, $D$ is the coefficient of DMI acting between the NNN spins in the linear regime of the spin-wave theory. $K_i$'s are the magnetic anisotropy constant, denoted as $K_A$, $K_B$, and $K_C$ for A, B, and C sublattice sites, respectively and acting along $z$-direction. The last term in Eq.~\eqref{Hamiltonian} is the Zeeman term, resulting from the external magnetic field $B_0$. Further, $\mb{e}_{ij}$ denotes the unit vector connecting $i^{\text{\tiny{th}}}$ to $j^{\text{\tiny{th}}}$ lattice sites and $\nu_{ij}$ is $+1(-1)$ for anticlockwise(clockwise) hopping from spin $i$ to spin $j$ as shown in Fig.~\ref{Dice}\tc{(b)}. 
The lattice vectors connecting A to B sublattice sites and C to A sublattice sites are denoted by $\delta_1 = a_0(0,1)$, $\delta_2 = a_0(\frac{\sqrt{3}}{2}, - \frac{1}{2})$ and $\delta_3 = a_0(-\frac{\sqrt{3}}{2}, - \frac{1}{2})$, whereas, the NNN vectors connecting two same sublattice sites are $\gamma_1=a_0(-\sqrt{3})$,  $\gamma_2 = a_0(\frac{\sqrt{3}}{2}, \frac{3}{2})$ and  $\gamma_3 = a_0(\frac{\sqrt{3}}{2}, -\frac{3}{2})$ respectively, as shown in Fig.~\ref{Dice}\tc{(c)}.

To study the low energy excitations or magnons for a ferromagnetic insulator described by Eq.~\eqref{Hamiltonian}, we have used linearised Holstein-Primakoff transformation~\cite{holstein1940field}, relating the spin operators ($\vec{S}$) and the bosonic operators ($a$, $a^\dagger$) at a site $i$,
\begin{equation}
        S_i^+ = \sqrt{2S}a_i; \hspace{5pt}
        S_i^- = \sqrt{2S}a_i^\dg; \hspace{5pt}
        S_i^z  = S - a_i^\dg a_i,
    \label{HP}
\end{equation}
where $S_i^+ = S_i^x + i S_i^y$, $S_i^- = S_i^x + i S_i^y $ and $a_i^\dg$($a_i$) are the creation(annihilation) operator at the $i^{\text{\tiny{th}}}$ position for A sublattice sites. For a ferromagnet, the same transformation will hold for B and C sublattice sites. The transformation renders Eq.~\eqref{Hamiltonian} to acquire a form,
\begingroup
\allowdisplaybreaks
\begin{align}
    H = & -JS\sum_{\braket{i,j}} \left( a_i^\dagger b_j + a_i b_j^\dagger 
        - a_i^\dagger a_i - b_j^\dagger b_j \right) \nonumber \\
      & - \frac{F}{2}S\sum_{\braket{i,j}} \Big[ \left( a_i^\dagger b_j + a_i b_j^\dagger \right) 
        + a_i^\dagger b_j^\dagger e^{i2\theta_\delta} + a_i b_j e^{-i2\theta_\delta} \Big] \nonumber \\
      & -JS\sum_{\braket{i,j}} \left( c_i^\dagger a_j + c_i a_j^\dagger 
        - c_i^\dagger c_i - a_j^\dagger a_j \right) \nonumber \\
      & - \frac{F}{2}S\sum_{\braket{i,j}} \Big[ \left( c_i^\dagger a_j + c_i a_j^\dagger \right) 
        + c_i^\dagger a_j^\dagger e^{i2\theta_\delta} + c_i a_j e^{-i2\theta_\delta} \Big] \nonumber \\
      & - S D\Big[\sum_{\braket{\braket{i,j}}\epsilon B} i\nu_{ij} b_i^\dagger b_j 
        + \sum_{\braket{\braket{i,j}}\epsilon C} i\nu_{ij} c_i^\dagger c_j + \text{H.C.}\Big] \nonumber \\
      & + \sum_i (K_A + \mu_B B_0) a_i^\dagger a_i 
        + \sum_i (K_B + \mu_B B_0) b_i^\dagger b_i \nonumber \\
      &~~~~~~~~~~~~~~~~+ \sum_i (K_C + \mu_B B_0) c_i^\dagger c_i,\label{real_Hamiltonian}
\end{align}
\endgroup
where $\theta_\delta$ is the angle between $\delta_i (i \in 1,2,3)$ and the $x$-direction.
Each site in the A sublattice of a dice geometry is a neighbor to both the B and C sublattice sites. While hopping from the $i^{\text{th}}$ to the $j^{\text{th}}$ spin, $\nu_{ij}$ takes the value $1(-1)$ depending on whether the hopping occurs via the B(C) sublattice sites, effectively canceling each other out, thereby prohibiting any hopping from A to A. Hence, the DMI is only effective for the B and C sublattice sites.

Now, for such a ferromagnetic dice model, the Fourier transformed Hamiltonian can be written in the Bogoliubov-de Gennes form, $\sum_k \psi^\dg(\mb{k}) \mathcal{H}(\mb{k})\psi(\mb{k})$, where $\psi(\mb{k})=(a_\mb{k}, a_{-\mb{k}}^\dg, b_\mb{k}, b_{-\mb{k}}^\dg, c_\mb{k}, c_{-\mb{k}}^\dg)^T$ and the explicit form of the matrix $\mathcal{H}(\mb{k})$ can be written as,
\begin{equation}
\footnotesize
     \begin{pmatrix}
                            M_A & 0 & - f(\mb{k}) & g^+(\mb{k}) & - f^*(\mb{k}) & g^{-*}(\mb{k})\\
                            0 & M_A & g^-(\mb{k}) & - f(\mb{k}) & g^{+*}(\mb{k}) & - f^*(\mb{k})\\
                            - f^*(\mb{k}) & g^{-*}(\mb{k}) & M_B-D' & 0 & 0 & 0\\
                            g^{+*}(\mb{k}) & - f^*(\mb{k}) & 0 & M_B-D' & 0 & 0\\
                            - f(\mb{k}) & g^+(\mb{k}) & 0 & 0 & M_C+D' & 0\\
                            g^-(\mb{k}) & - f(\mb{k}) & 0 & 0 & 0 & M_C+D'
                          \end{pmatrix},\label{matrix}
\end{equation}
where the onsite potentials are denoted by,
\begin{equation}
\small
    \begin{split}
        M_A &= K_A + \mu_B B_0 + 6 J,\\
        M_{B/C} &= K_{B/C} + \mu_B B_0 + 3 J,\\
        D'&= 2 D \sum_j \sin{\mb{k}\cdot \mb{\gamma}_j}
    \end{split}
\end{equation}
Further, the off-diagonal terms are written as,
\begin{equation}
\small
    f(k) = (J + \frac{F}{2})\sum_j e^{i k \delta_j},~g^{\pm}(k) = \frac{F}{2}\sum_j e^{\pm i 2 \theta_j} e^{i k \delta_j}.
\end{equation}
$\mathcal{H}(\mb{k})$ satisfies the generalized eigenvalue problem, $\eta \mathcal{H}(\mb{k})\mathcal{T}(\mb{k}) = \mathcal{T}(\mb{k}) \eta \mathcal{E}(\mb{k})$, where $\mathcal{T}(\mb{k})$ is the transformation matrix to diagonalize $\mathcal{H}(\mb{k})$, $\eta = I_{2 \times 2}\otimes \sigma_z$ is a metric and $\mathcal{E}(\mb{k})$ is an diagonal matrix whose elements are the eigenvalues of $\mathcal{H}(\mb{k})$. Now, to solve the eigenvalues of $\mathcal{H}(\mb{k})$ we introduce a non-Hermitian matrix $\mathcal{H'}(\mb{k}) = \eta \mathcal{H}(\mb{k})$, which has real spectra as long as the ferromagnetic state is an energy local minimum state~\cite{PhysRevLett.125.217202}. For bosons, only positive energy states are physical, hence, we shall focus solely on the positive energy spectrum in the following discussions. In the absence of DMI or PDI, the TRS is not violated, which, however, gets broken in their presence, implying $\mathcal{H}(-\mb{k}) = \mathcal{H}^*(\mb{k})$.

\section{RESULTS AND DISCUSSION}
\label{results}
It is important to realise that, in a dice lattice geometry, the A sublattice sites are connected to the B and C sublattice sites via NN exchange interactions. Consequently, the contribution to the onsite potential in the effective Hamiltonian is larger for the A sublattice sites than those for the other two as shown in Eq.~\eqref{matrix}. For convenience, we define $K_A = K_0$, $K_B = K_0+ \Delta$ and $K_C = K_0 - \Delta$, where, $K_0 = (K_A + K_B + K_C)/3$ and $\Delta = (K_B - K_C)/2$.
It is important to consider both $K_B > K_C$ and $K_C < K_B$ which can be achieved by considering both positive and negative values of $\Delta$. While the results should be symmetric with respect to $\Delta = 0$, however, the interplay with other parameters ($D$ and $F$) produces interesting results, which we shall elaborate on below.

At first, we look for the effects of both the DMI and the PDI on the nontrivial magnon band topology in the presence of a uniform magnetic anisotropy ($\Delta = 0$). However, to compare with the uniform scenario, we break the valley symmetry via $\Delta \neq 0$ and investigate the effects of nonuniform magnetic anisotropy on the magnon band topology. The rationale behind such a study is substantiated by various experimental and theoretical investigations. In experiments, the magnetic anisotropy can be tuned through various approaches, such as doping~\cite{yu2019tuning}, creating artificially layered systems~\cite{PhysRevMaterials.8.L011401}, etc. Additionally, from a theoretical perspective, the effects of the disorder have also been studied in magnetic systems, where the disorder induces significant changes in magnetic anisotropy and yields various topological properties~\cite{PhysRevLett.125.217202}. As we are dealing with a three-band model, it should be noted that in a three-band fermionic system, a pseudospin-1 model has been studied in the presence of a threefold degeneracy around the Dirac points. It has been shown that the presence of a mass term (in our system, denoted via $\Delta$) induces a coupling between the orbital magnetic moment and the external magnetic field, thereby leading to effective changes in both the longitudinal and Hall conductivity~\cite{PhysRevB.107.245150}. Thus, it should be possible to ascertain whether an analogous phenomenon persists in magnetic systems.

\subsection{Spectral Properties and Their Topological Aspects}
\label{uniform}
As emphasized earlier, a neat solution to the interacting spin model (Eq.~\eqref{Hamiltonian}) is possible via a spin-boson transformation. The Hamiltonian is subsequently solved with the DMI ($D$) and PDI ($F$) as parameters to introspect the topological properties of the magnon bands in a parameter space defined by these two. It is contextual to mention that these parameters are tunable in experiments~\cite{PhysRevLett.119.077205, belabbes2016oxygen}, while theoretically, it is of importance to assess their roles in shaping up the physical properties of interacting spin systems~\cite{dofhall, PhysRevB.105.214428}. 
\begin{figure}[b]
    \centering
    \includegraphics[width=0.9\columnwidth]{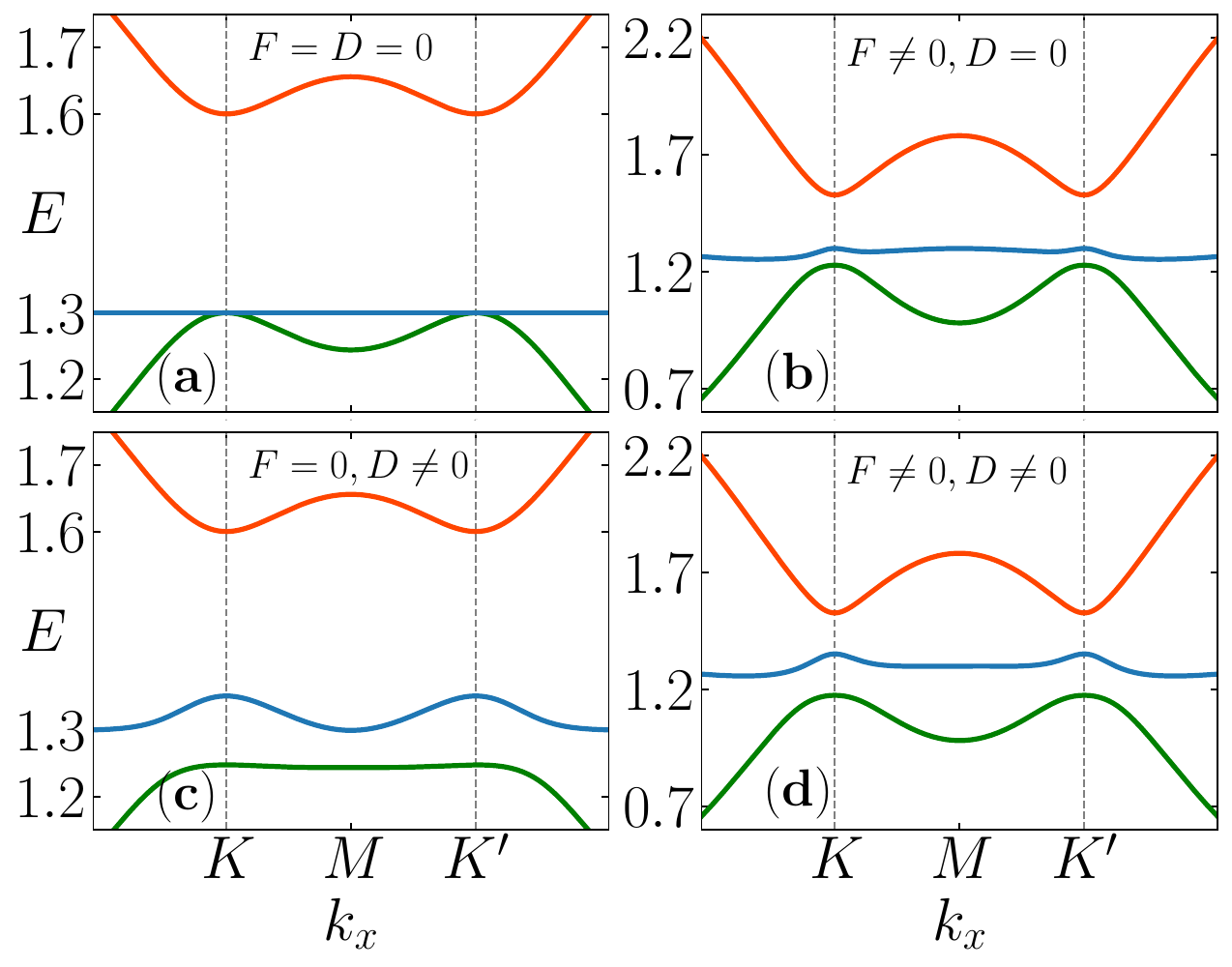}
    \caption{The bulk band structures are shown at $k_y=\frac{2\pi}{3}$. Here, the symmetry points are, $K = (-\frac{2\pi}{3\sqrt{3}}, \frac{2\pi}{3})$, $M = (0, \frac{2\pi}{3})$, $K'=(\frac{2\pi}{3\sqrt{3}}, \frac{2\pi}{3})$. The bulk band structures (a) without DMI and PDI, (b) in the presence of PDI ($F = 3J$) without DMI, (c) in the presence of DMI ($D = 0.1J$) without PDI, (d) in presence of both DMI ($D = 0.1J$) and PDI ($F = 3J$).}
    \label{Band_structure_1}
\end{figure}
Among the six bands in the dispersion (owing to the $6 \times 6$ matrix in Eq.~\eqref{Hamiltonian}), we consider only the three positive energy spectra, which we denote as the lower band (in green), the middle or the flat band (in blue), and the upper band (in red) in Fig.~\ref{Band_structure_1}~\cite{note}. In the absence of DMI and PDI, the lower band touches the flat band at the two Dirac points, namely at $K=(-\frac{2\pi}{3\sqrt{3}}, \frac{2\pi}{3})$ and $K'=(\frac{2\pi}{3\sqrt{3}}, \frac{2\pi}{3})$ as shown in Fig.~\ref{Band_structure_1}\tc{(a)} where we have kept $K_A = K_B = K_C$, that is, uniform magnetic anisotropy ($\Delta =0$). Throughout our investigation, we have kept the values of the parameters $K_0 = B_0 = 5 J$. However, due to the broken sublattice symmetry with the A sublattice at a higher onsite potential energy, the upper band remains gapped from the other two. Now, in presence of the PDI or the DMI, the flat band becomes distorted (we shall thus call it as the middle band), and is gapped from the lower band as shown in Figs.~\ref{Band_structure_1}\tc{(b)}, \tc{(c)}. Here, at the Dirac points, the energy difference between the upper and the middle bands is much larger than the difference between the middle and the lower bands. 

Considering both the interaction terms simultaneously, the upper band becomes relatively closer to the weakly dispersive middle band (see Fig.~\ref{Band_structure_1}\tc{(d)}). In Eq.~\eqref{real_Hamiltonian} the effective strength of the NN pseudodipolar interaction depends on the bond direction, which is eventually responsible for the weakly dispersive nature of the flat band. 

To focus on the topological properties, it needs to be ascertained whether the gap is trivial or topological. One of the ways to do that is by calculating the topological invariant, namely, the Chern number corresponding to those bands which is a valid invariant for TRS broken system. The Chern number ($C_n$) of a particular band can be calculated via~\cite{thouless1998topological},
\begin{equation}
    C_n = \frac{1}{2\pi} \int_{BZ} \Omega_n(\mb{k}) \cdot d\mb{k}, \label{Chern}
\end{equation}
where $n$ is the band index and $\Omega_n(\mb{k})$ is the Berry curvature corresponding to the $n^{\text{th}}$ band. Non-zero values of the Berry curvature arise due to the effects of broken TRS. The Berry curvature of the $n^{\text{th}}$ band is given by,
\begin{equation}
    \Omega^{xy}_n = -2\text{Im}\braket{\partial_{k_x}u_n(\mathbf{k})|\partial_{k_y}u_n(\mathbf{k})}. \label{Berry}
\end{equation}
Here $\ket{u_n(\mb{k})}$ is the eigenstate corresponding to the $n^{\text{th}}$ energy band of the Hamiltonian in Eq.~\eqref{Hamiltonian}. 

\begin{figure}[t]
    \centering
    \includegraphics[width=1\columnwidth]{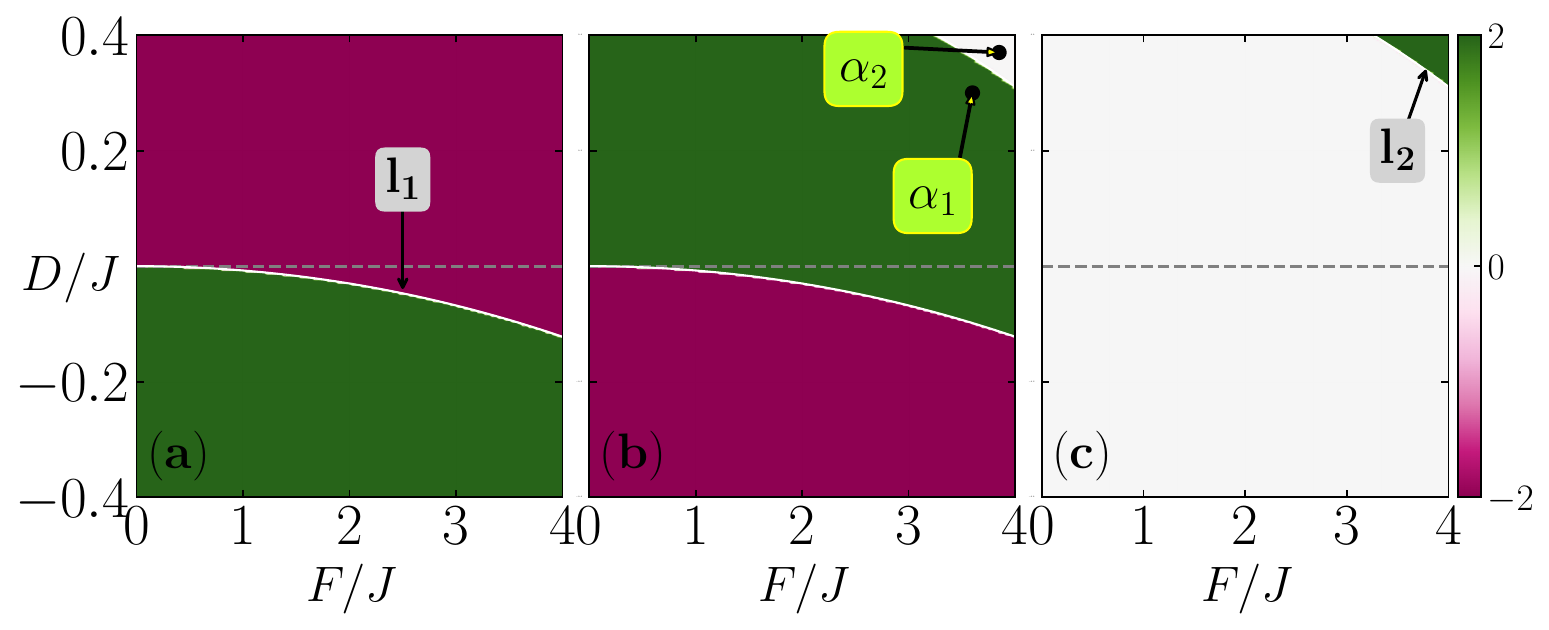}
    \caption{Phase diagrams are shown based on the values of the Chern number as a function of $F$ and $D$ in presence of uniform magnetic anisotropy. Phase diagrams corresponding to (a) the lower band, (b) the middle band, (c) the upper band are shown.}
    \label{phase_1}
\end{figure}

To demonstrate the interplay between the PDI and the DMI in the emergence of non-trivial phases and in inducing various topological phase transitions through multiple gap-closing scenarios between the three bands, we obtain the Chern number in the $F-D$ parameter space, as shown in the phase diagrams in Fig.~\ref{phase_1}. It may be noted that, although we consider $D$ to be either positive or negative depending on its direction, we consider only positive values of $F$. In a ferromagnet, $F$ is attractive, hence, negative values of $F$ has no physical significance.
In Figs.~\ref{phase_1}\tc{(a)} and \ref{phase_1}\tc{(b)}, we observe a phase transition from $C=+2 \rightarrow C=-2$ corresponding to the lower band and from $C = -2 \rightarrow C = +2$ for the middle band. This transition occurs across the curved line $l_1$, along which the middle and the lower bands touch each other at the Dirac points. By solving the Hamiltonian in Eq.~\eqref{Hamiltonian}, the transition line $l_1$ follows a mathematical relation, given by 
\begin{equation}
\small
\begin{split}
      M_C + 3\sqrt{3}D = \frac{1}{2}&\left[-3\sqrt{3}D - M_A + M_B \right.\\
                       &\left. + \sqrt{(M_A + M_B - 3\sqrt{3}D)^2 - 9F^2}  \right].   
\end{split}\label{l_1}
\end{equation}
In the absence of PDI ($F=0$), $l_1$ is along the line $D=0$. Further, we observe another phase transition occurring from a phase characterized by $C = +2 \rightarrow C = 0$ for the middle band and $C = 0 \rightarrow C=+2$ for the upper band, as shown in Fig.~\ref{phase_1}\tc{(b)} and \ref{phase_1}\tc{(c)}. This phase transition occurs due to the gap-closing between the middle and the upper bands, which occurs along the curved line $l_2$ and satisfies the relation 
\begin{equation}
\small
    \begin{split}
        M_C + 3\sqrt{3}D = \frac{1}{2}&\left[3\sqrt{3}D + M_A - M_C\right. \\
        &\left. + \sqrt{(M_A + M_B - 3\sqrt{3}D)^2 - 9F^2} \right]. 
    \end{split}\label{l_2}
\end{equation}

\begin{figure}[b]
    \centering
    \includegraphics[width=1\columnwidth]{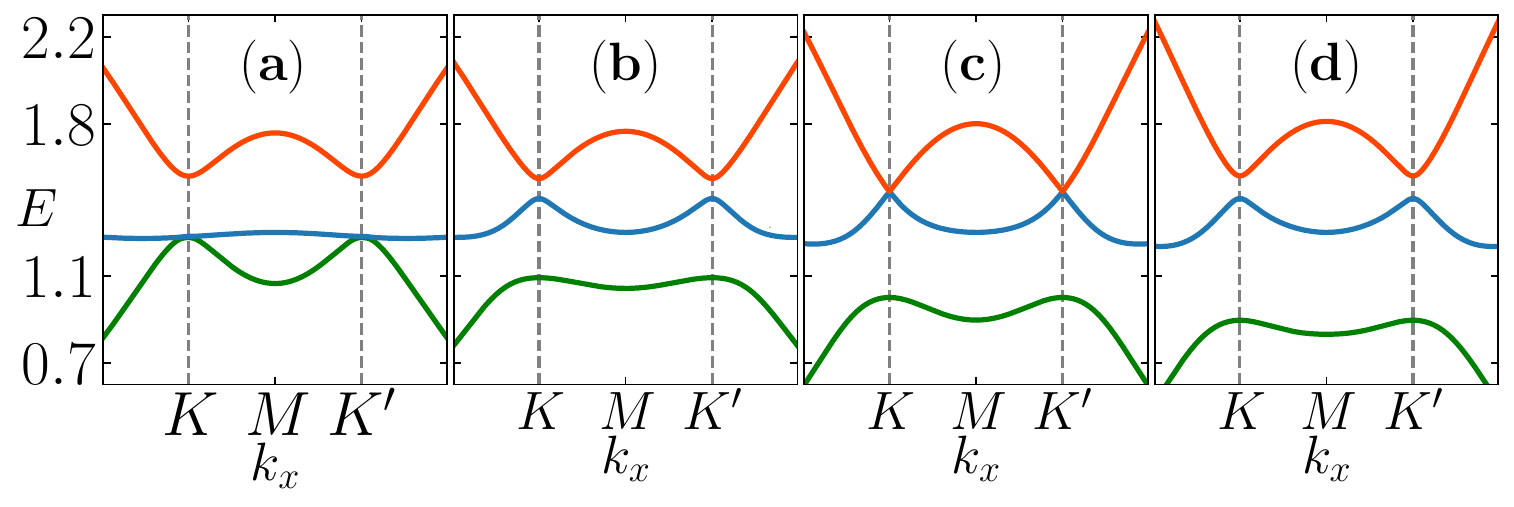}
    \caption{The bulk band structures are shown at $k_y=\frac{2\pi}{3}$. (a) One gap-closing event occurs when the parameters $F/J$ and $D/J$ lie on the $\text{l}_1$ line at $F = 2.298J$ and $ D = -0.034J$, (b) above which a gap opens at $F = 2.5J$ and $D = 0.3J$. (c) Another gap-closing phenomenon occurs when $F/J$ and $D/J$ lie on the $\text{l}_2$ line at $F = 3.6J$ and $D = 0.36J$, (d) and above that it reopens at $F = 4J$ and $D = 0.5J$. Here $\text{l}_1$ and $\text{l}_2$ lines refer to Fig.~\ref{phase_1}.}
    \label{Band_structure_2}
\end{figure}
To visualise the multiple topological phase transitions associated with different gap-closing scenarios, we have shown the bulk band spectra at those transition points. Along the line $l_1$ (see Eq.~\eqref{l_1}), despite finite values of both $F$ and $D$, the lower and middle bands touch each other in Fig.~\ref{Band_structure_2}\tc{(a)}, but they become gapped below and above $l_1$, as shown in Fig.~\ref{Band_structure_2}\tc{(b)}. Furthermore, if we increase the values of $F$ and $D$, at any point on the $l_2$ curve (see Eq.~\eqref{l_2}), the upper and middle bands touch each other at the Dirac points (both $K$ and $K'$ points), but they become gapped again for larger $F$ and $D$, as shown in Figs.~\ref{Band_structure_2}\tc{(c)} and \ref{Band_structure_2}\tc{(d)}.
So far, we have discussed how the flat band behaves in the presence of PDI and DMI under uniform magnetic anisotropy ($\Delta = 0$). Moving ahead, we shall explore the impact of nonuniform magnetic anisotropy ($\Delta \neq 0$) on the magnon band topology of a dice lattice in the subsequent analysis.

\begin{figure}[t]
    \centering
    \includegraphics[width=1\columnwidth]{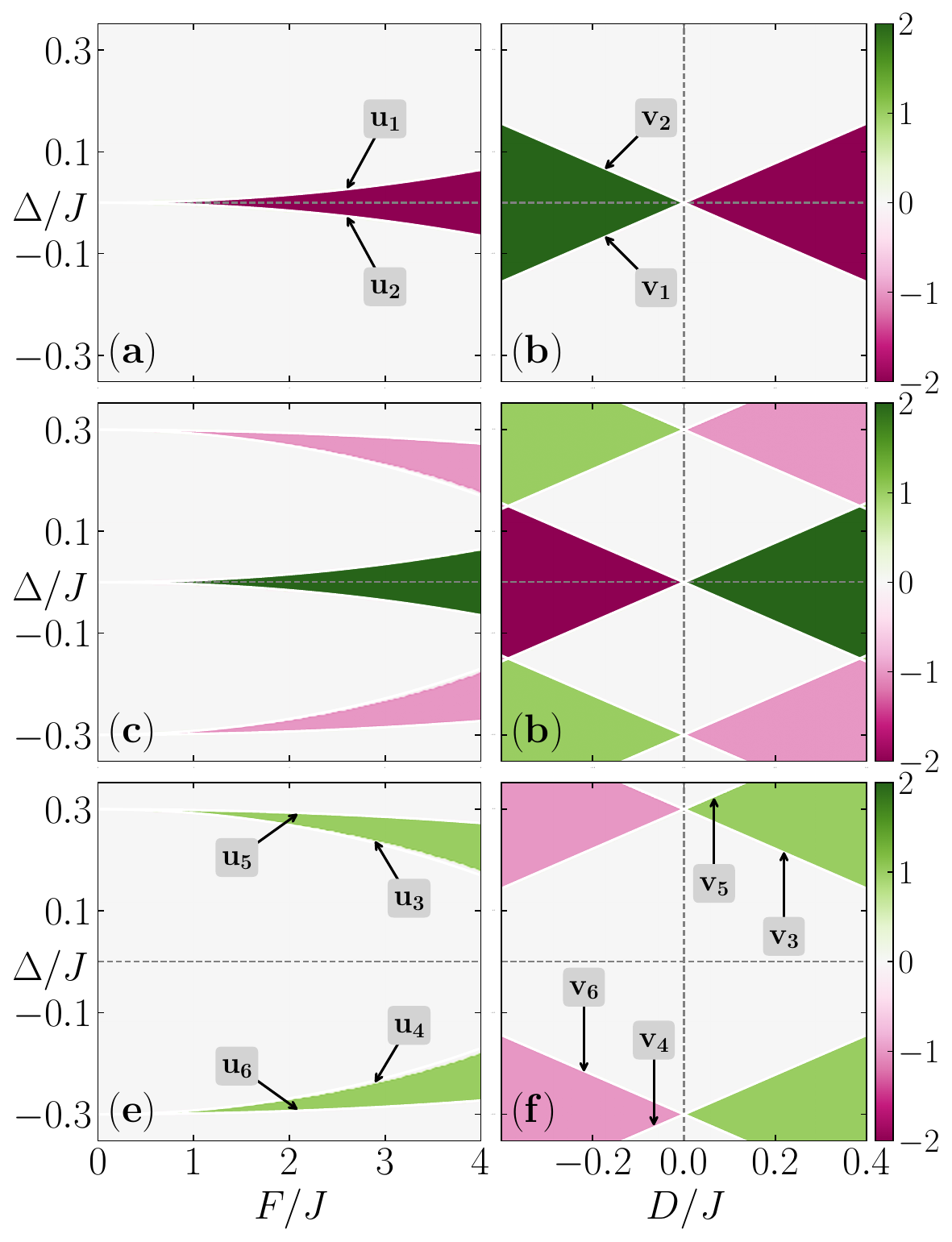}
    \caption{The phase diagrams based on the computation of the Chern number as a function of $\Delta/J$ and $F/J$ corresponding to (a) the lower band, (c) the middle band, and (e) the upper band are shown. Additionally, the phase diagrams are shown as a function of $\Delta/J$ and $D/J$ corresponding to (b) the lower band, (d) the middle band, and (f) the upper band are shown.}
    \label{phase-2}
\end{figure}

Under uniform magnetic anisotropy with $\Delta = 0$, and in the absence of PDI or DMI, the sublattice symmetry is broken due to the Heisenberg exchange interaction, leading to a trivial band gap, as depicted in Fig.~\ref{Band_structure_1}\tc{(a)}. By introducing a finite $\Delta$, the lower and the middle bands should gap out. However, this is insufficient to break the TRS, thereby resulting in a trivial magnon insulator. On the other hand, if there are TRS-breaking terms with $D \neq 0$ or $F \neq 0$ (or both), and no spatial inversion symmetry-breaking term, such as $\Delta$, the system exhibits nontrivial magnon topology. To further investigate the effect of a nonuniform magnetic anisotropy, we may obtain the Chern number as a function of $\Delta$ and look for its dependencies on $D$ and $F$ to construct a phase diagram as shown in Fig.~\ref{phase-2}. 

Here, we focus on the topological phase transitions in the $\Delta - F$  space in absence of DMI ($D = 0$). In Figs.~\ref{phase-2}\tc{(a)} and \ref{phase-2}\tc{(b)}, a phase transition occurs from $C=-2 \rightarrow C=0$ corresponding to the lower band and from $C = +2 \rightarrow C = 0$ for the middle band. The transitions follow the paths labeled as $\text{u}_1$ and $\text{u}_2$ in Fig.~\ref{phase-2}\tc{(a)}. Furthermore, $\text{u}_1$ and $\text{u}_2$ conform to the mathematical relation
\begin{equation}
\small
    \begin{split}
        M_C - \frac{1}{2}(-M_A + M_B + \sqrt{(M_A + M_B)^2 - 9F^2}) = \pm 2\Delta, 
    \end{split}\label{u_1u_2}
\end{equation}
where $2\Delta$ denotes the anisotropy difference between the $B$ and $C$ sublattices. Another phase transition is observed from $C = -1 \rightarrow C = 0$ for the middle band and from $C = +1 \rightarrow C = 0$ for the upper band. These transitions result from the gap-closing transitions occurring between the middle and the lower bands, and follow the transition lines $\text{u}_3$, $\text{u}_4$, $\text{u}_5$, and $\text{u}_6$, as depicted in Fig.~\ref{phase-2}\tc{(e)}. By solving the low-energy Hamiltonian $\mathcal{H}(\mb{k})$ at the vicinity of the Dirac points, we find that $\text{u}_3$ and $\text{u}_4$ adhere to the relation
\begin{equation}
\small
    \begin{split}
       M_B - \frac{1}{2}(M_A - M_B + \sqrt{(M_A + M_B)^2 - 9 F^2}) = \pm \Delta, 
    \end{split}\label{u_3u_4}
\end{equation}
while $\text{u}_5$ and $\text{u}_6$ satisfy the equation
\begin{equation}
    \small
    \begin{split}
        &M_C - \frac{1}{2}(-M_A + M_B + \sqrt{(M_A + M_B)^2 - 9 F^2})\\
        &~~=\pm [M_B - \frac{1}{2}(M_A - M_B + \sqrt{(M_A + M_B)^2 - 9 F^2})].
    \end{split}\label{u_5u_6}
\end{equation}
\begin{figure}[t]
    \centering
    \includegraphics[width=1\columnwidth]{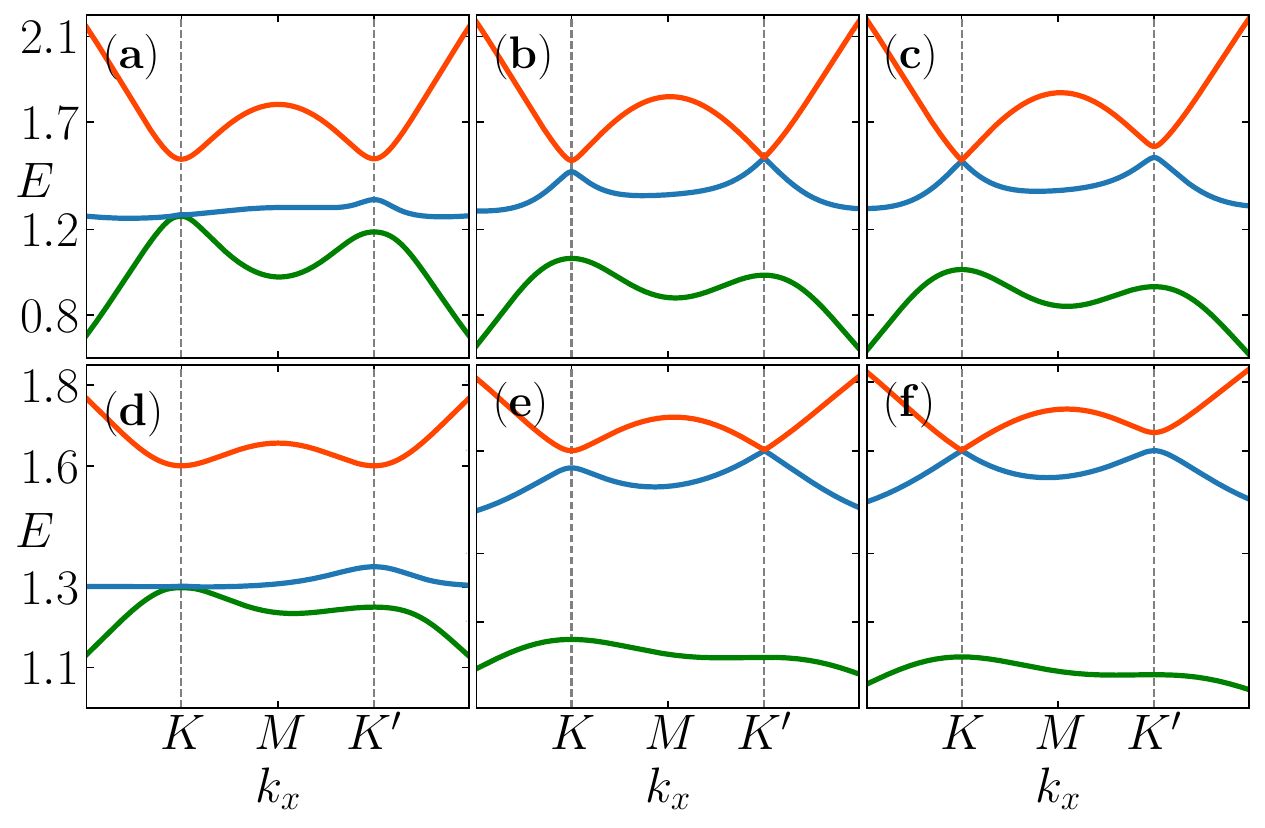}
    \caption{The bulk band structures are shown at $k_y=\frac{2\pi}{3}$. For $F=3J$ and $D = 0$, multiple gap-closing transitions occur as $\Delta$ is varied, when the parameters lie on (a) the $\text{u}_1$ line with $\Delta = 3.9J$, (b) $\text{u}_2$ line with $\Delta = 2.3J$ and (c) $\text{u}_3$ line with $\Delta = 2.87J$. Similarly, for $F=0$ and $D = 0.05J$, one observes multiple gap-closing transitions, when the parameters lie on the (d) $\text{v}_1$ line with $\Delta = 2.4J$, (e) $\text{v}_2$ line with $\Delta = 2.76J$ and (f) $\text{v}_3$ line with $\Delta = 3.27J$. Here $\text{u}_1$, $\text{u}_2$, $\text{u}_3$, $\text{v}_1$, $\text{v}_2$ and $\text{v}_3$ lines refer to Fig.~\ref{phase-2}.}
    \label{Band_structure_3}
\end{figure}

In continuation of a similar exercise, we investigate the topological behavior of the system in the presence of nonuniform magnetic anisotropy and DMI, without any pseudodipolar interaction or PDI. As an NNN interaction term, the DMI is considered much weaker than the PDI and can be either positive or negative depending on its hopping orientation ($\nu_{ij}$). In Figs.~\ref{phase-2}\tc{(b)} and \ref{phase-2}\tc{(d)}, a topological-to-trivial phase transition is observed from $C = \mp2 \rightarrow C = 0$ for the lower band and from $C = \pm2 \rightarrow C = 0$ corresponding to the middle band. For any value of $D$, $\Delta$ being tuned both for positive and negative values leads to these phase transitions just across the $\text{v}_1$ and $\text{v}_2$ lines (see Fig.~\ref{phase-2}\tc{(b)}). The analytical expression governing $\text{v}_1$ and $\text{v}_2$ is
\begin{equation}
    \small
    \Delta = \pm 3\sqrt{3}D. \label{v_1v_2}
\end{equation}

As $\Delta$ is further enhanced, an additional phase transition occurs from $C=0$ to $C=\pm 1$ and to $C=0$ for the lower band, and from $C=0$ to $C=\mp 1$ and finally to $C=0$ for the upper band, as shown in Figs.~\ref{phase-2}\tc{(d)} and \ref{phase-2}\tc{(f)}. These phase transitions are accompanied by gap-closing and reopening scenarios between the middle and the upper bands, and follow the paths $\text{v}_3$, $\text{v}_4$, $\text{v}_5$, and $\text{v}_6$ depicted in Fig.\ref{phase-2}\tc{(f)}. The paths $\text{v}_3$ and $\text{v}_4$ adhere to the relation,
\begin{equation}
    \small
    (3J - 3\sqrt{3}D) = \pm \Delta,\label{v_3v_4}
\end{equation}
 while $\text{v}_5$ and $\text{v}_6$ follow the relation 
 \begin{equation}
     \small
     (3J + 3\sqrt{3}D) = \pm \Delta. \label{v_5v_6}
 \end{equation}

All the phase transitions mentioned above occur through gap-closing transitions, which may be discernible in the bulk band structure. Hence, we show the bulk band properties at the high symmetry points ($K$, $M$, and $K'$) corresponding to the phase transitions. In the absence of DMI, when $F$ and $\Delta$ satisfy the mathematical relation for $\text{u}_1$, given in Eq.~\eqref{u_1u_2}, the lower and the middle bands touch at the $K$ point, in Fig.~\ref{Band_structure_3}\tc{(a)}. Further, if we increase $\Delta$, while keeping $F$ fixed, and the parameters satisfy the relation corresponding to $\text{u}_3$ and $\text{u}_5$ (Eqs.~(\ref{u_3u_4},~\ref{u_5u_6})), we get two additional gap-closing transitions at the $K'$ and $K$ points respectively, where the middle and the upper bands are involved as shown in Figs.~\ref{Band_structure_3}\tc{(b)},~\ref{Band_structure_3}\tc{(c)}. Moreover, if we consider negative values of $\Delta$, one will observe similar gap-closing transitions at $K$ and $K'$ points satisfying the mathematical relations corresponding to $\text{u}_2$, $\text{u}_4$, $\text{u}_6$ (Eqs.~(\ref{u_1u_2},~\ref{u_3u_4},~\ref{u_5u_6})). Similarly, if we consider variation of $D$ and $\Delta$ in the absence of PDI, and the parameters satisfy the relation for $\text{v}_1$ (see Eq.~\eqref{v_1v_2}), the computation of the Chern number reveals a gap-closing event occurring at the $K$ point between the lower and middle bands, as shown in Fig.\ref{Band_structure_3}\tc{(d)}. By further increasing $\Delta$ while keeping $D$ constant, two additional gap-closing events occur, one each at the $K'$ and $K$ points, satisfying the transition lines $\text{v}_3$ and $\text{v}_5$ respectively (see Eqs.~(\ref{v_3v_4},~\ref{v_5v_6})), involving the upper and the middle bands, as depicted in Figs.\ref{Band_structure_3}\tc{(e)} and \ref{Band_structure_3}\tc{(f)}. The same transitions occur when the parameters satisfy the relation given in Eqs.~(\ref{v_1v_2},~\ref{v_3v_4},~\ref{v_5v_6}) corresponding to $\text{v}_2$, $\text{v}_4$, $\text{v}_6$, for negative values of $\Delta$. However, it is important to note that if a gap-closing phenomenon occurs at the $K$ point for a positive value of $D$, the corresponding gap-closing event will occur at the $K'$ point for a negative value of $D$, and vice versa. Hence, we get distinct topological phases and phase transitions therein having the same magnitude, corresponding to positive and negative values of $D$ (Fig.~\ref{phase-2}).

\begin{figure}[b]
    \centering
    \includegraphics[width=1\columnwidth]{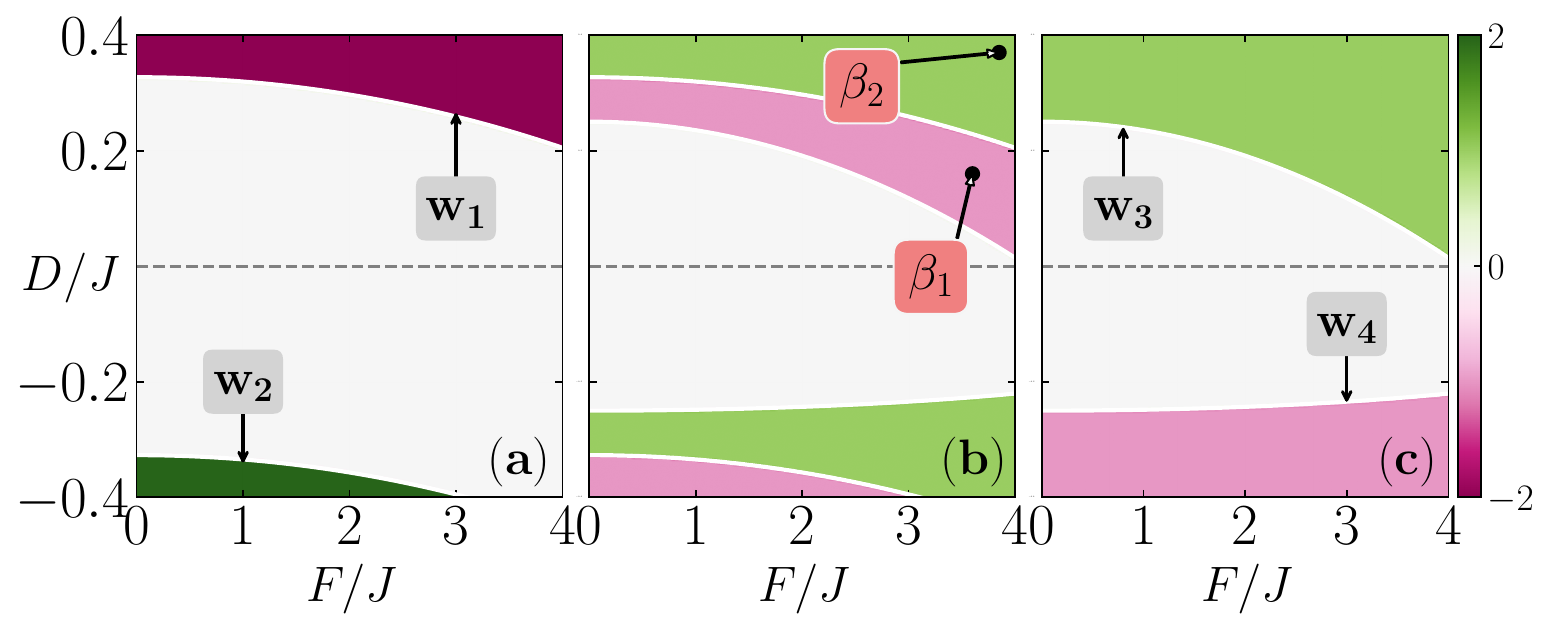}
    \caption{The phase diagrams are shown based on the values of the Chern number as a function of $F$ and $D$, under nonuniform magnetic anisotropy at $\Delta = 1.7J$. The phase diagrams corresponding to (a) the lower band, (b) the middle band, and (c) the upper band are shown.}
    \label{phase-3}
\end{figure}
So far, we have discussed how the DMI or the PDI individually influences the topological properties of the system under nonuniform magnetic anisotropy. For completeness, and to examine their properties, we shall explore the topological behavior of the system when both the interactions are present, with nonuniform magnetic anisotropy ($\Delta \neq 0$). Hence, we have calculated the Chern numbers corresponding to the three bands in the $F$ and $D$ space defined with a particular value of $\Delta$, namely, $\Delta = 1.7J$ as shown in Fig.~\ref{phase-3}.

\begin{figure}[b]
    \centering
    \includegraphics[width=1\columnwidth]{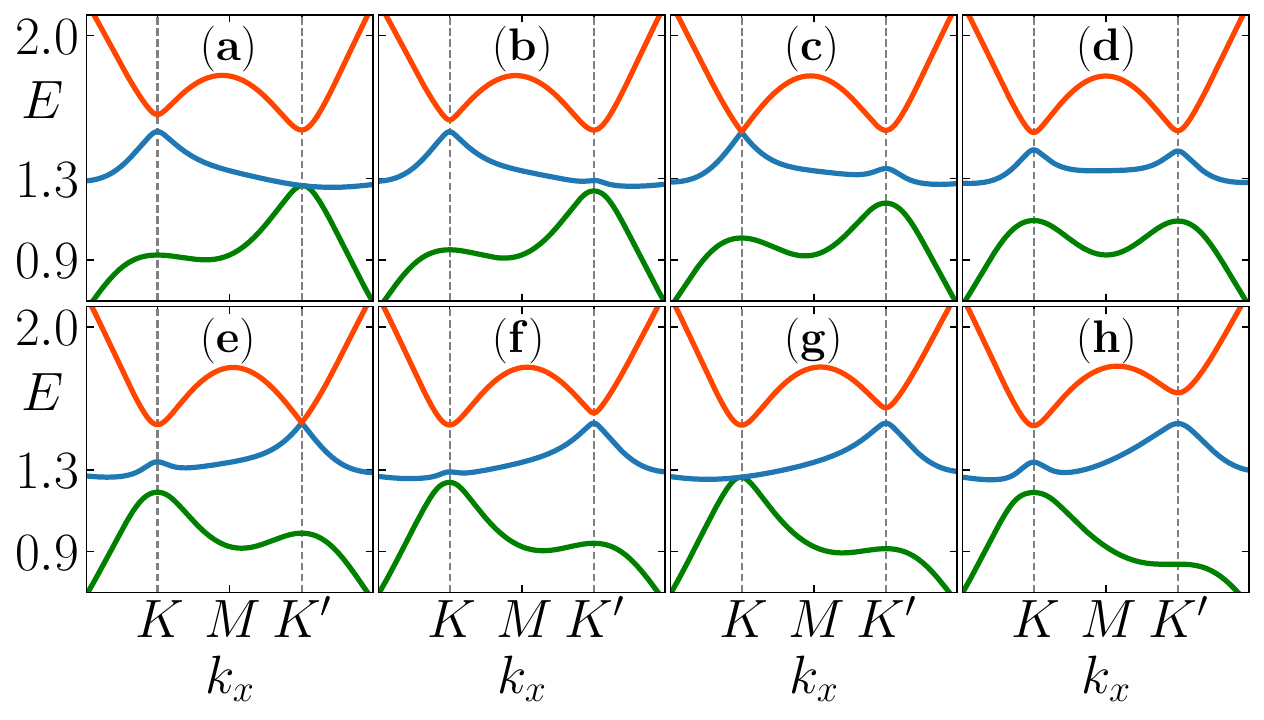}
    \caption{The bulk band structures are shown at $k_y=\frac{2\pi}{3}$. At $F=3J$, $\Delta = 1.7J$ we vary $D$ from $-0.4J$ to $0.4J$. (a) one gap-closing transition occurs when the parameters lie on the $\text{w}_2$ line, (b) beyond that the gap becomes opened. (c) Another gap-closing transition occurs when they cross the $\text{w}_4$ line, and (d) the bands become gapped beyond that, indicating a trivial phase. (e) Upon varying $D$ gives rise to another gap-closing along the $\text{w}_3$ line, and (f) beyond that, the bands become gapped. (g) Another phase transition occurs by gap-closing when the parameters satisfy $\text{w}_1$ line and (h) finally gapped out beyond that. Here $\text{w}_1$, $\text{w}_2$, $\text{w}_3$ and $\text{w}_4$ lines refer to Fig.~\ref{phase-3}.}
    \label{Band_structure_4}
\end{figure}
In Fig.~\ref{phase-3}, multiple topological phase transitions are observed. Keeping $F$ fixed, we ascertain the dependence on the DMI term for positive values of $D$, which demonstrates a topological phase transition between the middle and the upper bands, characterized by $C = 0 \rightarrow C=\mp 1$, that occurs above the transition line $\text{w}_3$ (see Figs.~\ref{phase-3}\tc{(b)},~\ref{phase-3}\tc{(c)}). At $\text{w}_3$, the parameters satisfy the relation,
\begin{equation}
    \small
    \begin{split}
       M_B + 3\sqrt{3}D = \frac{1}{2}&\left[3\sqrt{3}D + M_A - M_B\right.\\
       &\left. + \sqrt{(M_A + M_B - 3\sqrt{3}D)^2 - 9F^2}\right]. 
    \end{split}\label{w_3}
\end{equation}
 Beyond this transition, if $D$ is further enhanced, we observe another phase transition above $\text{w}_1$ from $C = 0 \rightarrow C=-2$ for the lower band, and from $C=-1 \rightarrow C=+1$ for the middle band (see Figs.~\ref{phase-3}\tc{(a)},~\ref{phase-3}\tc{(b)}). At $\text{w}_1$, the parameters obey the relation,
 \begin{equation}
     \small
     \begin{split}
        M_C + 3\sqrt{3}D = \frac{1}{2}&\left[-3\sqrt{3}D - M_A + M_B \right.\\
        &\left. + \sqrt{(M_A + M_B - 3\sqrt{3}D)^2 - 9F^2}\right]. 
     \end{split}\label{w_1}
 \end{equation} 
In contrast for $D$ being negative, two distinct topological phase transitions are observed. As the parameters cross the transition line $\text{w}_4$, given by 
\begin{equation}
    \small
    \begin{split}
        &-3\sqrt{3}D - M_A + M_B + \sqrt{(M_A + M_B - 3\sqrt{3}D)^2 - 9F^2}\\ 
        &~~~= (3\sqrt{3}D + M_A - M_C + \sqrt{(M_A + M_C - 3\sqrt{3}D)^2 - 9F^2}),
    \end{split}\label{w_4}
\end{equation}
 one phase transition occurs between the middle and the upper bands, from $C = 0 \rightarrow C=\pm 1$. Beyond this transition, when the value of $D$ crosses the transition line $\text{w}_2$, given by
 \begin{equation}
     \small
     \begin{split}
         M_B + 3\sqrt{3}D = \frac{1}{2}&\left[3\sqrt{3}D + M_A - M_C \right.\\
         &\left.+ \sqrt{(M_A + M_C - 3\sqrt{3}D)^2 - 9F^2}\right],
     \end{split}\label{w_2}
 \end{equation}
another phase transition occurs from $C=0 \rightarrow C=+2$ for the lower band and from $C=-1 \rightarrow C=+1$ for the middle band. 

To add further depth into our discussion, in Fig.~\ref{Band_structure_4} we have shown all the gap-closing and reopening phenomena through the bulk band structures of our system. Here we consider a specific value of PDI, namely, $F=3 J$, and vary $D$ in the range $-0.4J$ to $0.4J$. One phase transition occurs when $D$ crosses the $\text{w}_2$ line, the lower and middle bands touch each other at the $K'$ point, and beyond that, the gap reopens, as shown in Figs.~\ref{Band_structure_4}\tc{(a)},~\ref{Band_structure_4}\tc{(b)}. However, the system remains topological with non-zero values of the Chern numbers. If we further increase $D$, when the parameters satisfy the relation corresponding to the $\text{w}_4$ transition line (Eq.~\eqref{w_4}), we observe another gap-closing event among the middle and the upper band touching each other at the $K$ point. Beyond that, the system becomes trivial with zero Chern number and gapped bands, in Figs.~\ref{Band_structure_4}\tc{(c)},~\ref{Band_structure_4}\tc{(d)}. The system becomes again topological when $D$ crosses the $\text{w}_3$ line (Eq.~\eqref{w_3}), where the gap between the upper and the middle bands closes at $K'$ point, and becomes gapped again beyond the transition (see Figs.~\ref{Band_structure_4}\tc{(e)},~\ref{Band_structure_4}\tc{(f)}). Another gap-closing event occurs when the parameters satisfy the relation corresponding to $\text{w}_1$ transition line (Eq.~\eqref{w_1}), where the lower and the middle bands touch each other at the $K$ point and resurface that transition for larger values of $D$ (see Figs.~\ref{Band_structure_4}\tc{(g)},~\ref{Band_structure_4}\tc{(h)}). Again, the system continues to possess its topological properties.

\subsection{Magnonic Edge Modes}
\label{edge_modes}
\begin{figure}[b]
    \centering
    \includegraphics[width=1\columnwidth]{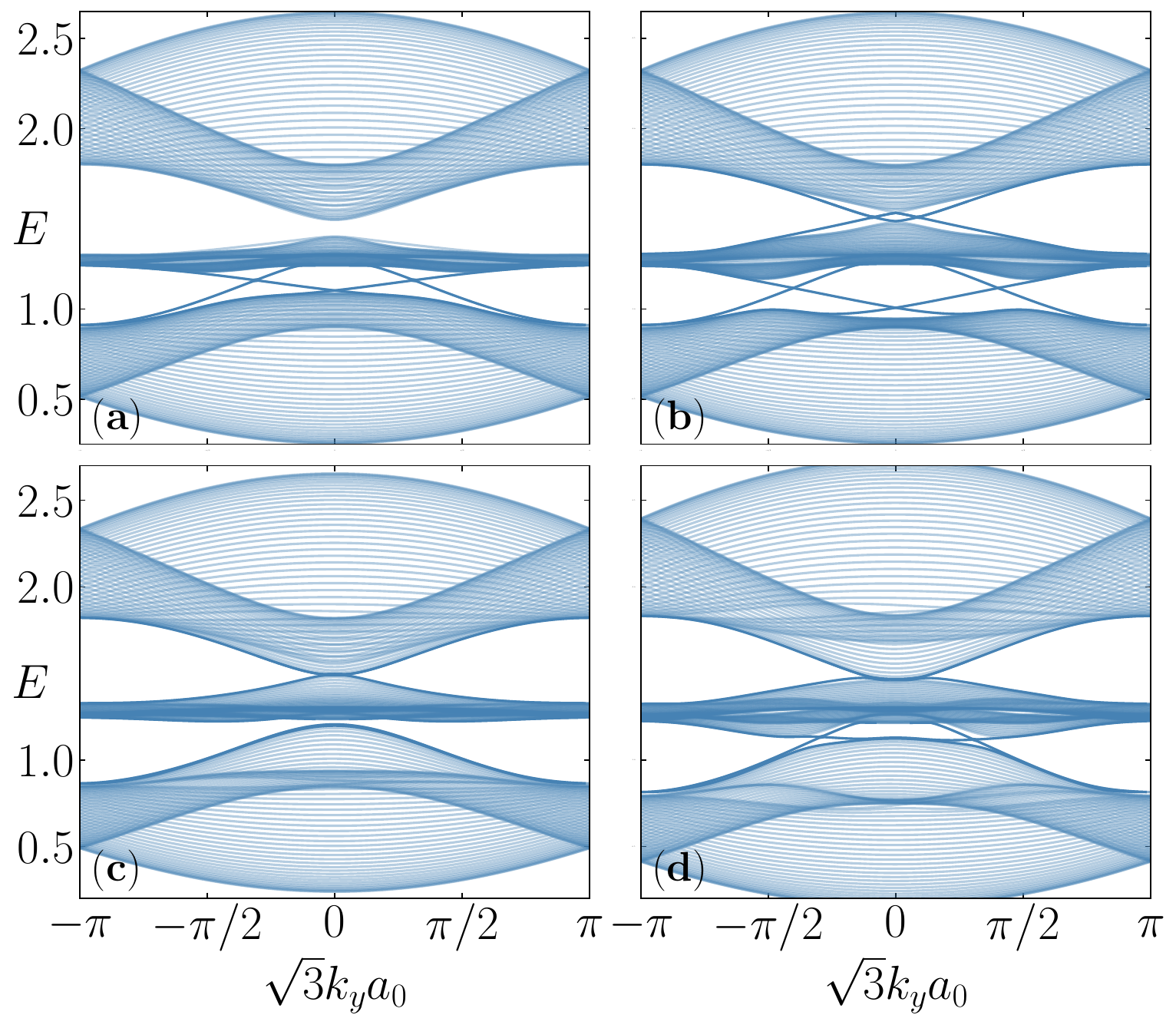}
    \caption{The bulk boundary correspondence has been shown at four distinct topological phases, which are marked by (a) $\alpha_1$, and (b) $\alpha_2$ in Fig.~\ref{phase_1} and (c) $\beta_1$, and (d) $\beta_2$ in Fig.~\ref{phase-3}. The calculations are performed in an armchair nanoribbon with 60 unit cells along $x$-direction.}
    \label{ribbon}
\end{figure}
Further, to investigate the magnonic edge modes or the bulk boundary correspondence, we consider a semi-infinite ribbon, where the system has a finite number of lattice sites in one direction, while maintaining periodic boundary conditions in the other direction. The number of crossings of the magnonic edge states at the $n^{\text{th}}$ band gap is the sum of the Chern numbers up to $n^{\text{th}}$ bands. Also, the sign of the sum denotes the direction of propagation of the magnonic edge modes. The sum of the Chern numbers up to $n^{\text{th}}$ band gap is also known as the winding number~\cite{mook2014edge}, given by,
\begin{equation}
    \mathcal{W}_n = \sum_{i = 0}^n C_i, \label{winding}
\end{equation}
where $i$ is the band index. We take a ribbon with an armchair edge to maintain open boundary conditions with finite lattice sites (60 unit cells) along $x$-direction and periodic boundary conditions with infinite lattice sites along $y$-direction. Hence, we can Fourier transform Eq.~\eqref{real_Hamiltonian} along the $y$-direction, considering $k_y$ as a good quantum number which yields a set of six coupled eigenvalue equations presented in Appendix~\ref{appendix:Edge modes}.

We have computed $\mathcal{W}_n$ using Eq.~\eqref{winding} to characterize distinct topological phases both in the presence of uniform and nonuniform magnetic anisotropy. For the uniform case, we show the bulk boundary correspondence at two distinct topological phases, denoted by $\alpha_1$ and $\alpha_2$ in Fig.~\ref{phase_1}. Here, prior to the gap-closing transition between the upper and the middle bands, the Chern numbers associated with the bands, from bottom to top, have values $-2$, $2$, and $0$ respectively. Consequently, as mentioned in Eq.~\eqref{winding}, two edge crossings occur in the gapped region between the middle and the lower bands ($\mathcal{W}_1 = -2$), while no such edge crossing is observed in the upper gapped region ($\mathcal{W}_2 = 0$), as shown in Fig.~\ref{ribbon}\tc{(a)}. Beyond a gap-closing transition, the Chern numbers change to $-2$, $0$, and $2$ obtained corresponding to the bottom to the top bands respectively. As a result, in Fig.~\ref{ribbon}\tc{(b)}, two edge crossings emerge in each of the gapped regions; one between the middle and the lower bands ($\mathcal{W}_1 = -2$), and another between the upper and middle bands ($\mathcal{W}_2 = -2$).

A similar scenario is considered in the presence of nonuniform magnetic anisotropy, where the broken valley symmetry reveals the system to undergo multiple topological phase transitions. Hence, we choose two distinct topological phases, named as $\beta_1$ and $\beta_2$ as depicted in Fig.~\ref{phase-3}. At $\beta_1$, only the upper and the middle bands correspond to non-zero Chern numbers $C = \pm 1$, whereas the Chern number corresponding to the lower band is zero. Hence, no edge modes are observed between the middle and the lower bands ($\mathcal{W}_1 = 0$) as shown in Fig.~\ref{ribbon}\tc{(c)}. However, at $\beta_2$, all the bands acquire non-zero Chern numbers, denoted by $C = -2$, $+1$, $+1$ respectively, moving from bottom to top. Hence, we observe two edge modes between the lower and the middle bands ($\mathcal{W}_1 = -2$), where the other two bands touch each other ($\mathcal{W}_2 = -1$) in the ribbon structure (Fig.~\ref{ribbon}\tc{(d)}). 
The negative sign in $\mathcal{W}_n$ ($n = 1,2$) indicates that the magnon modes propagate from right to left. However, in phase space, for negative values of $D$, a positive Chern number is obtained for the lower bands, and in this case, the magnonic edge modes traverse from left to right.

\subsection{Thermal Hall Conductivity}
\label{thermal}
\begin{figure}[t]
    \centering
    \includegraphics[width=1\columnwidth]{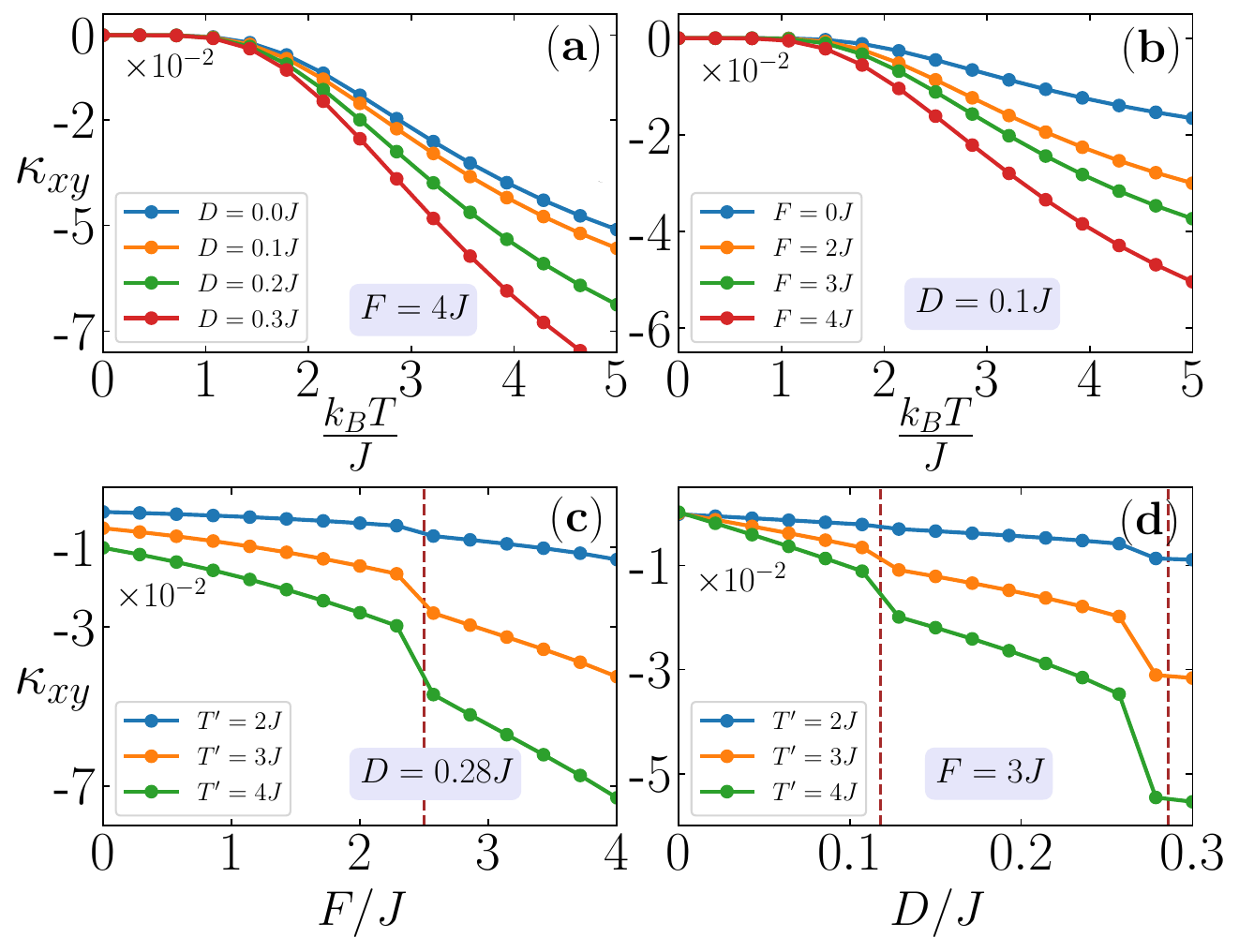}
    \caption{The behaviour of the thermal Hall conductivity as a function
            of temperature (a) at $F=4J$ as a function of $D$ and (b) at $D=0.1J$ varing $F$. The behaviour of the thermal Hall conductivity (c) as a function of $F$ at $D= 0.28J$ and as a function of $D$ at $F = 3J$ at three different temperatures. The red dashed line shows the phase transitions as obtained from the phase diagram parallel in Fig.~\ref{phase-3}.}
    \label{Thermal_Hall}
\end{figure}

We have explicitly studied the topological invariant and the edge modes for our system. We can calculate the transport properties based on the band structures. In fermionic systems, the electrons occupy all the states up to the Fermi level, leading to quantized Hall conductivity. However, in bosonic systems, with no such concept of the Fermi energy, all states contribute to the transport properties, including the Hall conductivity. Moreover,  at low temperatures, all the particles intend to be in the lowest energetic eigenstate. Hence, the Hall conductivity receives a major contribution from lower energy states. The thermal Hall conductivity can be expressed as~\cite{thermalhall},
\begin{figure}[b]
    \centering
    \includegraphics[width=1\columnwidth]{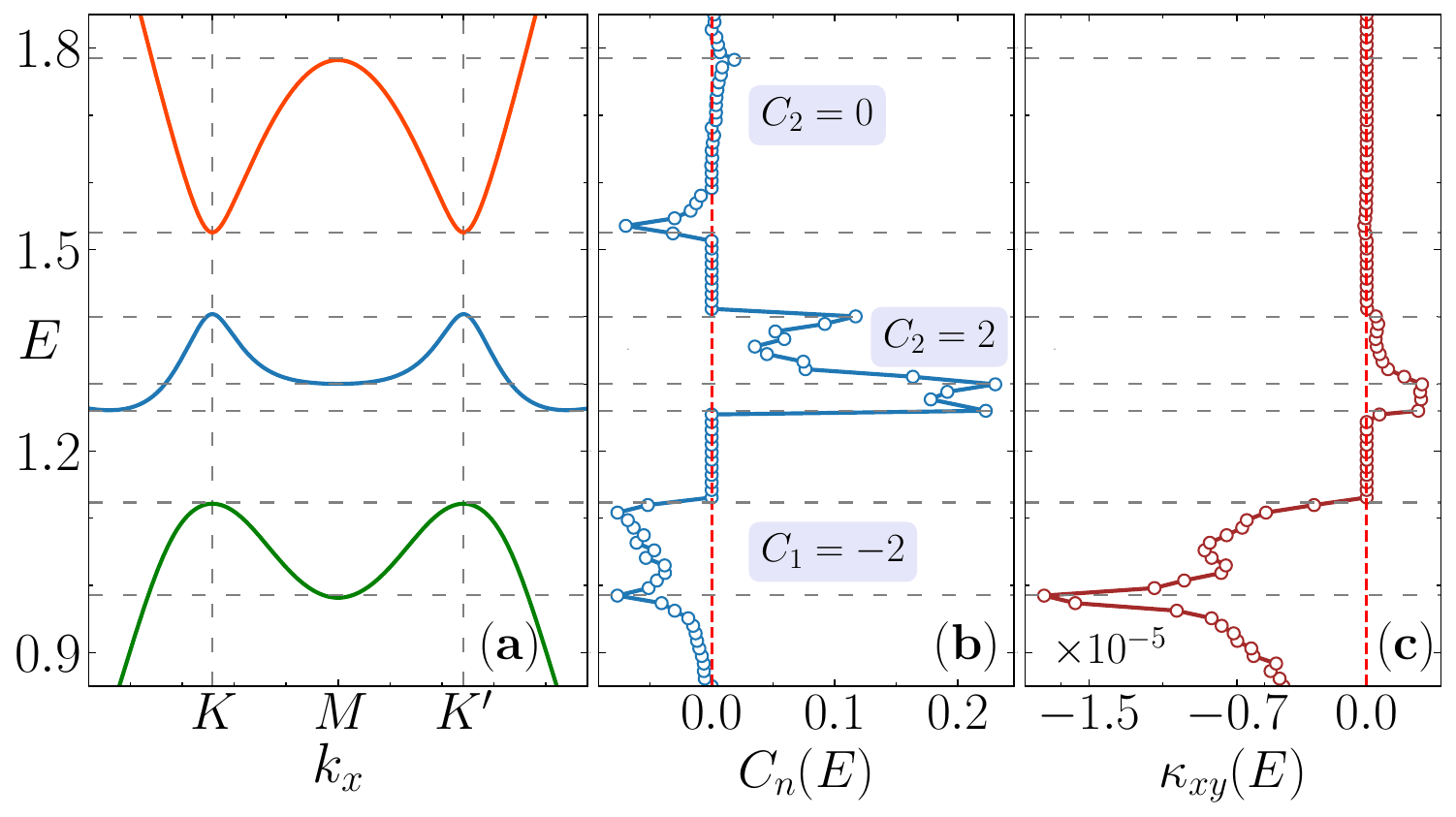}
    \caption{The behaviour of the energy-dependent Chern number and the thermal Hall conductivity as a function
            of the band energy are shown. (a) The bulk band structures at $F=4J$, $D=0.2J$, $\Delta= 0$ are shown. (b) Corresponding energy-dependent Chern number and (C) the energy-dependent thermal Hall conductivity at $T = J$.}
    \label{Chern_Hall}
\end{figure}
\begin{equation}
    \kappa_{xy} = \frac{k_B T}{4 \pi^2} \sum_n \int_{BZ} c_2(\rho_{n,\mathbf{k}})\Omega_{xy}^n(\mathbf{k}) \cdot d\mb{k}, \label{kappa}
\end{equation}
where $T$ denotes the absolute temperature, $\rho_{n,\mathbf{k}}$ is the Bose-Einstein distribution function, given by $\rho_{n,\mathbf{k}} = 1 / [\exp(\varepsilon_n(\mathbf{k})/k_BT)-1]$, $\varepsilon_n(\mathbf{k})$ is the energy eigenvalue corresponding to the $n^{\text{th}}$ eigenstate, and $c_2(\rho_{n,\mathbf{k}})$ is the weighting function with 
\begin{equation}
\begin{split}
    c_2(\rho_{n,\mathbf{k}})&=(1 + \rho_{n,\mathbf{k}})\left(\log{\frac{1 + \rho_{n,\mathbf{k}}}{\rho_{n,\mathbf{k}}}}\right)^2- (\log{x})^2 \\
    &~~~~~~~~~~~~~~~~~~~~~~~- 2\text{Li}_2 (- \rho_{n,\mathbf{k}}), 
\end{split}\label{c2}
\end{equation}
where $\text{Li}_2(-\rho_{n,\mathbf{k}})$ is a polylogarithmic function~\cite{Li2}. 
To arrive at a comprehensive understanding of how the thermal Hall conductivity is influenced by the magnon band structure with a view to establish a close connection between them, we introduce Chern numbers $C_n(E)$ corresponding to the isoenergy surfaces with energy $E$~\cite{PhysRevB.89.134409}, given by,
\begin{equation}
    C_n(E) = \frac{1}{2\pi}\int_{BZ}\delta(E - E_n)\Omega_{xy}^n(\mathbf{k}) \cdot d\mb{k}.\label{Chern_energy}
\end{equation}
The corresponding energy-dependent thermal Hall conductivity is given by,
\begin{equation}
    \kappa_{xy}(E) = \frac{k_B T}{(2\pi)^2} \sum_n \int_{BZ} \delta(E - E_n) c_2(\rho_{n,\mathbf{k}})\Omega_{xy}^n(\mathbf{k}) \cdot d\mb{k}, \label{kappa_energy}
\end{equation}
where the delta function incorporates the restriction of using an isoenergy surface and $n$ is the band index. In Fig.~\ref{Chern_Hall}, we observe the behavior of the energy-dependent Chern number and thermal Hall conductivity in relation to the band structure. In general, the Berry curvature is concentrated near the high symmetry points, where the bands exhibit higher curvature. Hence, $C_n(E)$ obtained via integrating the Berry curvature over the BZ, becomes larger at extremum energies as shown in Fig.~\ref{Chern_Hall}\tc{(b)}. On the other hand, along with the Berry curvature, the thermal Hall conductivity also contains a weight function $c_2(\rho_{n,\mathbf{k}})$ in the integrand, which is higher for low energies~\cite{cao2015magnon}, indicating that the primary contribution to the thermal Hall conductivity arises from the low energy spectra. Hence, $\kappa_{xy}(E)$ decreases with increasing energy (see Fig.~\ref{Chern_Hall}\tc{(c)}). Moreover, the sign of the Hall conductivity reflects the value of topological invariant associated with the lower band, as shown in Fig.~\ref{Thermal_Hall}.
 
Considering the parameter ranges of the Hamiltonian in this theoretical investigation, the Heisenberg exchange interaction strength $J$ is assumed to be much smaller than the other interaction energies, except for the DMI. In the presence of a small $J$, to stabilize the ferromagnetic state under strong spin-orbit coupling, we consider the presence of a sufficiently high external magnetic field along with an easy-axis magnetic anisotropy energy ($B = K = 5J$). Hence, we vary the temperature from zero to $5J$, which pertains to a reasonable temperature regime for observing the thermal Hall effect in our system.
With a uniform magnetic anisotropy, in Fig.~\ref{Thermal_Hall}\tc{(a)}, we have shown in the presence of the PDI, the thermal Hall conductivity, $\kappa_{xy}$ monotonically increases as a function of temperature and the DMI strength $D$. Similarly, in the presence of DMI, $\kappa_{xy}$ exhibits a similar monotonic increase as a function of temperature ($T$) and the PDI strength $F$, in Fig.~\ref{Thermal_Hall}\tc{(b)}. Thus, in the absence of any discontinuity, no phase transition occurs, thereby supporting the corresponding phase plots presented in Fig.~\ref{phase_1}. If $T$ is further increased, at very high temperature ($T\rightarrow \infty$), $\kappa_{xy}$ will saturate at a finite value, known as $\kappa_{xy}^{lim}$~\cite{PhysRevB.89.134409}. Although $\kappa_{xy}^{lim}$ reflects the strength of the magnon Hall conductivity, it cannot be obtained within a ferromagnetically ordered phase.

Further, to reflect on the phase transition occurring as an interplay of $D$ and $F$ in presence of nonuniform magnetic anisotropy, for a fixed value of $D$, in Fig.~\ref{Thermal_Hall}\tc{(c)}, we observe a discontinuity in the thermal Hall conductivity $\kappa_{xy}$ near $F = 2.5J$ as $F$ is varied, which support the phase transition depicted in Fig.~\ref{phase-3}\tc{(b)}. Moreover, we observe two such discontinuities in $\kappa_{xy}$, as a function of $D$ at a fixed value of $F=0.3$, one near $D \simeq 0.28J$, another close to a value  $D \simeq 0.12J$, as shown in Fig.~\ref{Thermal_Hall}\tc{(d)}.
Further, at low temperatures, the function $c_2(\rho_{n,\mathbf{k}})$ (see Eq.~\eqref{c2}) exhibits a more rapid variation as a function of $E$ compared to its behaviour at higher temperatures~\cite{PhysRevB.89.134409}. Higher values of either $F$ or $D$, or both reduce the energy of the lower band, thereby making $c_2(\rho_{n,\mathbf{k}})$ more dominant than the contribution arising from the Berry curvature in influencing the behaviour of the observed thermal Hall conductivity. Consequently, in Figs.~\ref{Thermal_Hall}\tc{(c)} and \ref{Thermal_Hall}\tc{(d)}, the jumps in $\kappa_{xy}$ at the onset of phase transitions are less apparent at low temperatures compared to that at higher temperatures.
\begin{figure}[t]
     \begin{subfigure}[!h]{0.375\columnwidth}
         \includegraphics[width=\columnwidth]{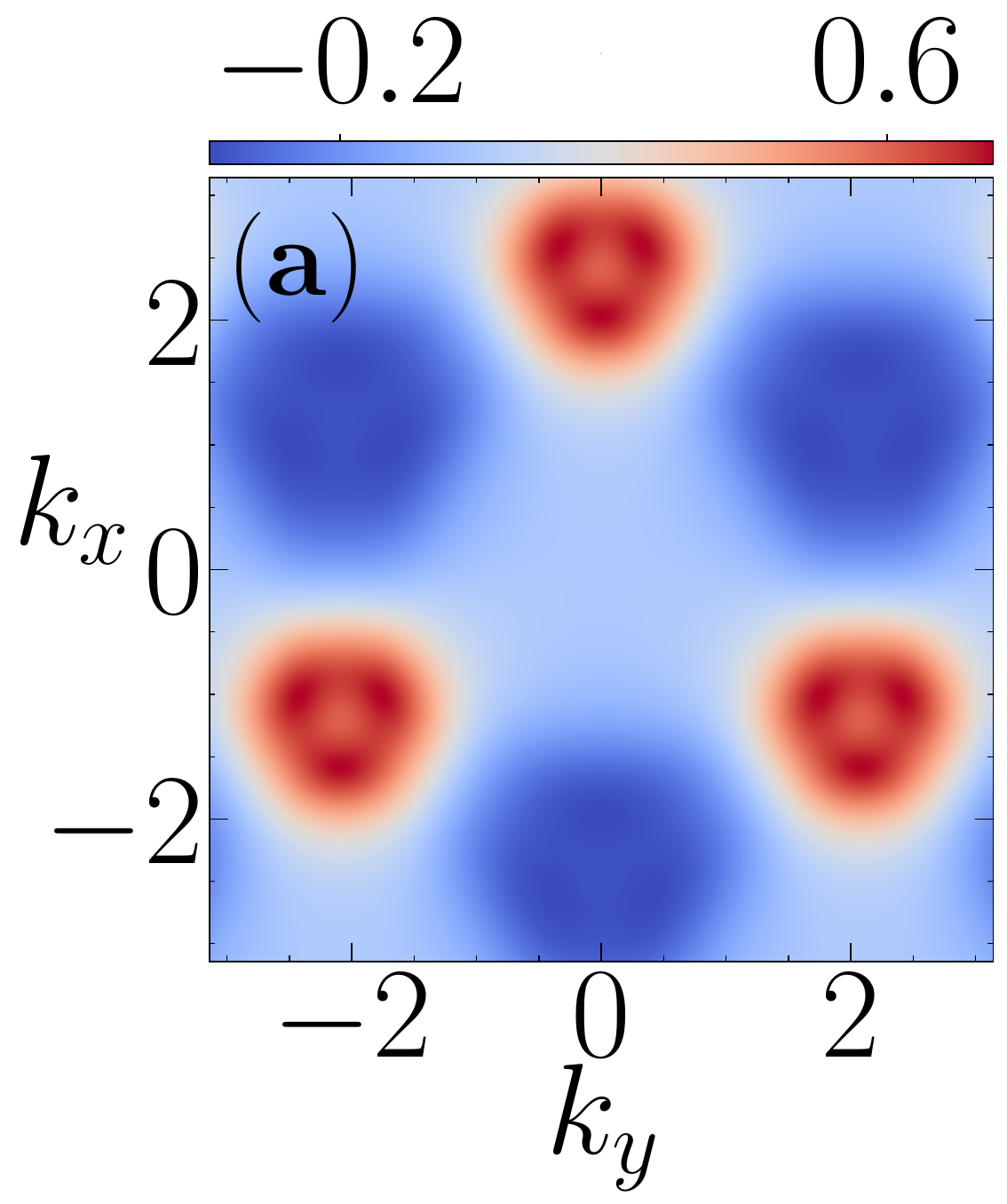}
         \captionlistentry{}
     \end{subfigure}
     \hspace{-5 pt}
     \begin{subfigure}[!h]{0.3025\columnwidth}
         \includegraphics[width=\columnwidth]{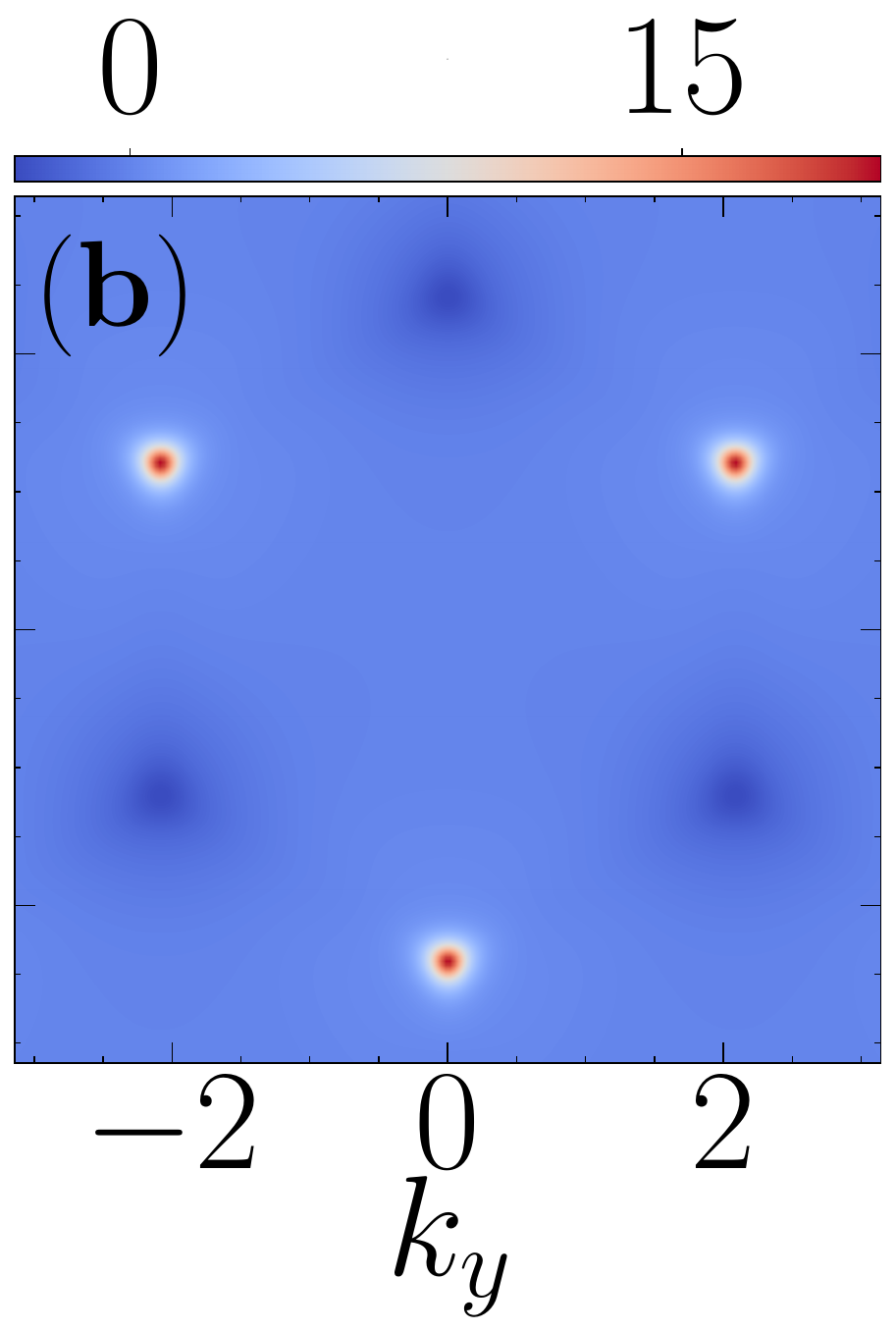}
         \captionlistentry{}
     \end{subfigure}
     \hspace{-5 pt}
     \begin{subfigure}[!h]{0.3025\columnwidth}
         \includegraphics[width=\columnwidth]{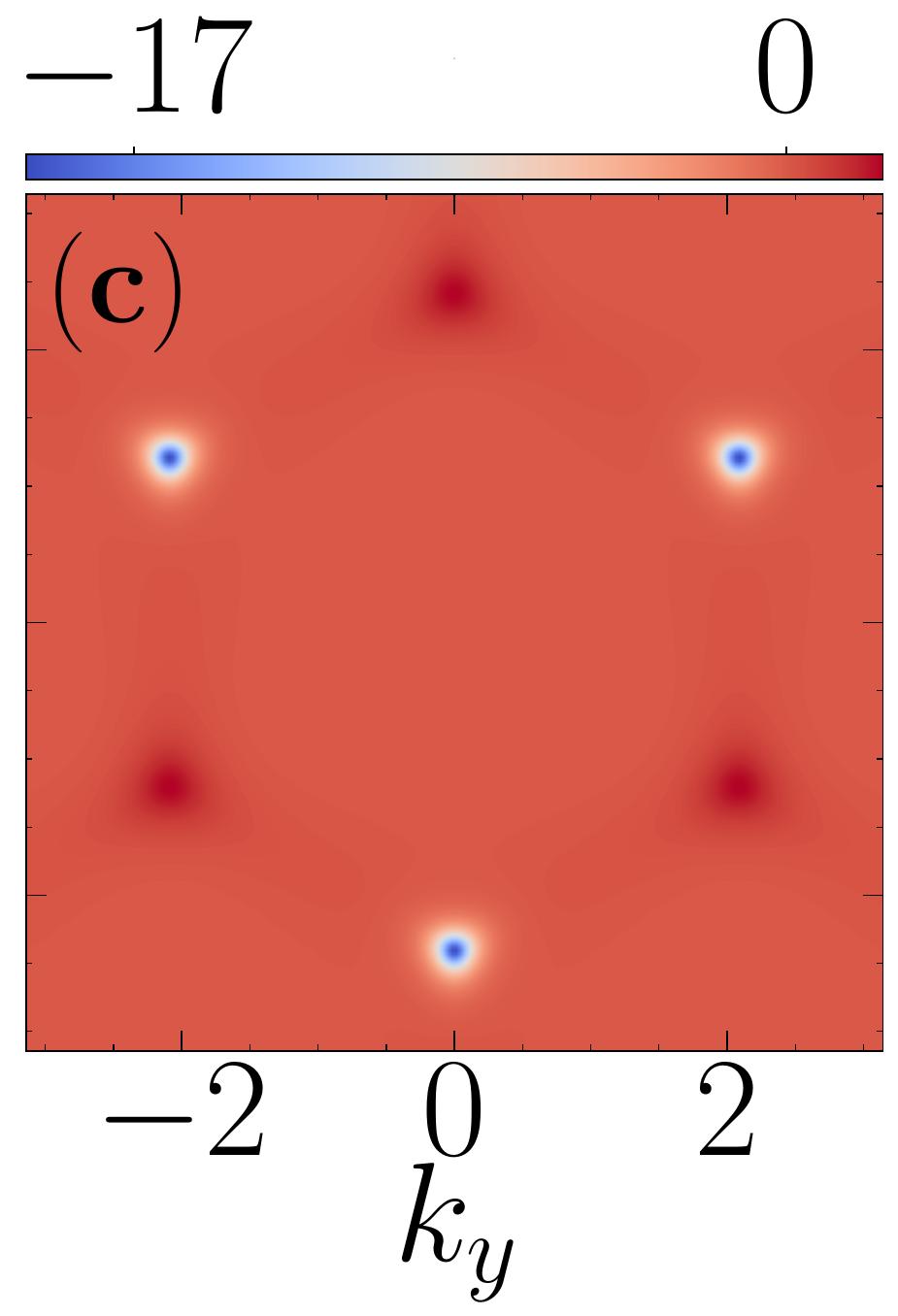}
         \captionlistentry{}
     \end{subfigure}
     \caption{Non-zero Berry curvature for (a) the lower band, (b) the middle band, (c) the upper band at the trivial region are shown.}
     \label{Berry_plot}
\end{figure}

Further, according to the phase diagram in Fig.~\ref{phase-3}, at $F=3J$ and $-0.25J < D< 0.12J$, the Chern numbers corresponding to all the bands are zero. However, in presence of both the PDI and the DMI, TRS is broken in the system, which causes a non-zero Berry curvature as shown in Fig.~\ref{Berry_plot}, which is eventually responsible for the non-zero Hall conductivity (Fig.~\ref{Thermal_Hall}\tc{(d)}).

Furthermore, It is important to note that, in this theoretical investigation, we have shown that the thermal Hall conductivity increases monotonically with temperature. However, experimental studies reveal that, due to thermal fluctuations, $\kappa_{xy}$ begins to decrease beyond a certain temperature~\cite{kim2024thermal}, which has a value usually between $\sim 2J$ to $\sim 3J$. Although a higher external magnetic field can partially diminish the effects of thermal fluctuations~\cite{PhysRevB.85.134411}, it is generally safer to study the Hall response at low temperatures in real experiments.

\section{HONEYCOMB LATTICE: SIMPLIFIED KITAEV MODEL}
\label{Honeycomb_Lattice}
\begin{figure}[b]
    \centering
    \includegraphics[width=0.9\columnwidth]{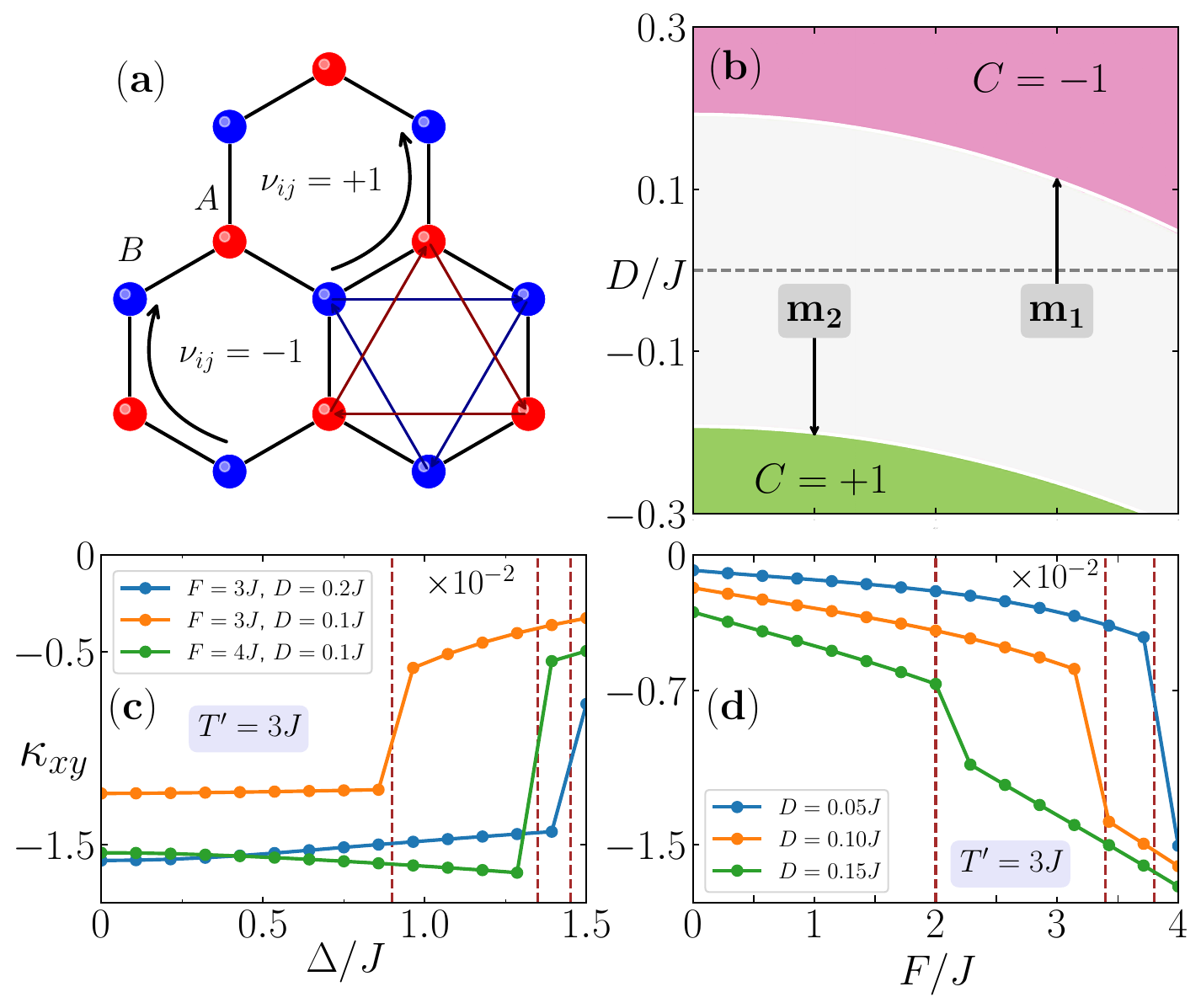}
    \caption{(a) Schematic diagram for a honeycomb lattice and (b) the phase diagram corresponding to the lower band as a function of $D$ and $F$ with nonuniform magnetic anisotropy with $\Delta = J$. The behaviour of the thermal Hall conductivity as a function of (c) $\Delta$ and (d) $F$.}
    \label{Honeycomb_results}
\end{figure}

For the sake of completeness, it is important to address the corresponding results for a simple honeycomb lattice, that is without a central atom, as shown in Fig.~\ref{Honeycomb_results}\tc{(a)}. The resulting Hamiltonian is
\begin{equation}
    \begin{split}
        &H = - J \sum_{\braket{i,j}}\left(\mb{S}_i \cdot \mb{S}_j\right) - F\sum_{\braket{i,j}}\left(\mb{S}_i\cdot \mb{e}_{ij}\right)\left(\mb{S}_j\cdot \mb{e}_{ij}\right) \\
        & + D\sum_{\braket{\braket{i,j}}\in A}\nu_{ij}\hat{z}\left(\mb{S}_i \times \mb{S}_j\right) + D\sum_{\braket{\braket{i,j}}\in B}\nu_{ij}\hat{z}\left(\mb{S}_i \times \mb{S}_j\right) \\
        &-\frac{1}{2}\sum_{i \in A} K_A S_{zi}^2 -\frac{1}{2}\sum_{i \in B} K_B S_{zi}^2 - \mu_B B_0 \sum_i S_{zi},
    \end{split}\label{Hamiltonian_honeycomb}
\end{equation}
 where all the symbols have similar significance as that of Eq.~\eqref{real_Hamiltonian} with only two sublattice sites, namely, A and B. The Hamiltonian in Eq.~\eqref{Hamiltonian_honeycomb} can be mapped onto the Kitaev spin model~\cite{kitaev2006anyons}. Applying the linearised Holstein-Primakoff transformation as described via Eq.~\eqref{HP}, one obtains the effective (bosonic) Hamiltonian. In the presence of the PDI or the DMI, spectral properties of this two band model have been studied earlier in the literature. However, to provide an essential comparison with the magnons on a dice lattice, here we discuss the interplay of DMI and the PDI in presence of nonuniform magnetic anisotropy ($K_A\neq K_B$), with $K_0 = (K_A + K_B)/2$ and $\Delta = (K_A - K_B)/2$.

To introduce nonuniform magnetic anisotropy, we set a particular value, namely, $\Delta = J$. Under these conditions, a topological phase transition is observed as a function of $F$ and $D$ (in unit of $J$), as shown in Fig.~\ref{Honeycomb_results}\tc{(b)}. While in a dice lattice with various magnetic interactions, the topological phase can exhibit a higher Chern number, namely, $C=\pm 2$, in the absence of a third (flat or dispersive) band, the topological phases in the honeycomb lattice are characterized by a Chern number of $C = \pm 1$. That is why all the topological phases and the phase transitions therein exhibit more exotic features in a dice lattice compared to those in a honeycomb monolayer. 

Here, in Fig.~\ref{Honeycomb_results}\tc{(c)}, we observe a phase transition from a topological to a trivial and returns to a topological phase ($C=\pm 1 \rightarrow C=0 \rightarrow C=\mp 1$). For a fixed $F$, the dependency on $D$ can be ascertained. We observe a transition as $D$ crosses the $\text{m}_2$ line (see Fig.~\ref{Honeycomb_results}). Beyond this point, the system exhibits a trivial phase. Further, at some positive value of $D$, near the $\text{m}_1$ line, we observe another phase transition from $0$ to $-1$, beyond which the system remains topological. The mathematical relation corresponding to $\text{m}_1$ and $\text{m}_2$ lines are given by,
\begin{equation}
    \small
        (M'_B + 3 \sqrt{3} D) - \frac{1}{2}(\sqrt{(M'_A + M'_B - 6 \sqrt{3} D)^2 - 9 F^2}) = \pm 2 \Delta,
        \label{m_1m_2}
\end{equation}
where $M'_A = M'_B = K_0 + B_0 + 3J$.

Similar to the dice lattice, all the phase transitions occur through gap-closing phenomena at the high-symmetry points $K$ or $K'$. While this phase diagram corresponds to the lower band of the magnon spectrum, a similar phase diagram can be observed for the upper band, where the sign of the Chern number is simply reversed (that is $C= \pm 1 \rightarrow C=\mp 1$). 

As previously mentioned, owing to the distribution function $c_2(\rho_{n,\mb{k}})$ in Eq.~\eqref{kappa}, we observe the thermal Hall conductivity to track all the phases and the phase transitions corresponding to the lower band. In Fig.~\ref{Honeycomb_results}\tc{(c)}, the thermal Hall conductivity in the presence of the PDI or the DMI, is impacted by the disparity in magnetic anisotropy between the two sublattices, namely A and B. We observe a sudden jump, at $\Delta = \Delta_c$, where the system becomes trivial. Higher values of $F$ or $D$ require larger $\Delta_c$ for the phase transition to occur. The red dotted lines indicate the phase transition and the corresponding value of $\Delta_c$. Although we should expect nearly zero thermal Hall conductivity beyond the phase transition, a very small finite Hall conductivity is observed due to the broken time-reversal symmetry. Further, in Fig.~\ref{Honeycomb_results}\tc{(d)}, as a function of  $F$, we observe the thermal Hall conductivity at three distinct values of $D$, replicating and hence supporting the phase transition observed in Fig.~\ref{Honeycomb_results}\tc{(b)}. 

While the system described in Eq.\eqref{Hamiltonian_honeycomb} has a lower Chern number as compared to that of the dice system, we observe relatively lower values for the thermal Hall conductivity in Fig.~\ref{Honeycomb_results}\tc{(c)} and \tc{(d)}. This highlights the motivation for studying flat band (or multiband) systems with higher Chern numbers, which are expected to yield larger thermal Hall conductivity and hence, easily attainable in experiments. However, unlike fermionic systems, where the Hall conductivity is quantized in unit of $\frac{e^2}{h}$, magnonic systems lack an equivalent quantized Hall effect.

\section{CONCLUSION}
\label{conclusion}
We have shown that nontrivial band topology can be observed in the dice lattice magnetic system with ferromagnetic exchange interaction in the presence of magnetic anisotropy, DMI, PDI, or all of them. Our motivation was to ascertain the interplay of their effects in inducing topological phase transitions and obtaining the thermal Hall effect. Although the effects of PDI and DMI have been studied individually in literature, in the presence of high spin-orbit coupling, both interactions can arise in the system. The band structures and the phase plots reveal the existence of distinct topological phases, with multiple topological phase transitions resulting from several gap-closing and reopening events at the high symmetry Dirac points. 

Let us enumerate the highlights of our findings. Under uniform magnetic anisotropy, the sublattice symmetry is broken via the onsite potential of the $A$ sublattice due to the Heisenberg exchange interaction. There, we observe topological phase transitions as a function of $F$ and $D$. The topological phases have also been replicated via the magnonic edge modes obtained in the semi-infinite ribbon band dispersions. Further, under nonuniform magnetic anisotropy, where additionally the spatial inversion symmetry or valley symmetry is broken between $B$ and $C$ sublattices, we observe various topological to trivial phase transitions regardless of the system with persisting broken time-reversal symmetry. In both the cases, we have explicitly shown all the mathematical relations followed by the gap-closing events or the phase transitions. Owing to the presence of a higher number of bands, expectedly we have higher Chern numbers than a two-band model. Transport studies in the form of thermal Hall conductivity further replicate the topological invariants obtained in our work. Additionally, this conductivity is capable of capturing all relevant phase transitions in presence of nonuniform magnetic anisotropy. Furthermore, for the sake of completeness and to highlight the effects of the flat band, we compare the results for a dice lattice with those for a honeycomb structure in the presence of all the magnetic interactions discussed above, which denotes specialized Kitaev honeycomb model with uniform exchange interactions.

At a fundamental level, our results focus on the topological behaviors and nontrivial transport properties of a multiband magnetic system in the presence of DMI or PDI.

\appendix 
\section{Eigenvalue Equations For Edge Modes}
\label{appendix:Edge modes}
The six coupled eigenvalue equation for 60 unit cells along $x$-direction obtained from Eq.~\eqref{real_Hamiltonian} is given by
\begin{subequations}
 \begingroup
 \allowdisplaybreaks
    \begin{align}
        E_k &a_{k,n} = -(J + F/2) \lbrace ( b_{k,n} e^{i\eta_1 k} + c_{k,n} e^{-i\eta_2 k} )  \nonumber \\
        &+( b_{k,n+1} + c_{k,n+1} ) + ( b_{k,n-1} + c_{k,n-1} ) \rbrace -\nonumber \\
        &(F/2)\lbrace ( b_{-k,n}^\dagger e^{i\eta_1 k} + c_{-k,n}^\dagger e^{-i\eta_2 k}) e^{i2\theta_1} \nonumber\\
        &+ \left( b_{-k,n+1}^\dagger + c_{-k,n-1}^\dagger \right) e^{i2\theta_2}\nonumber\\
        &+ ( b_{-k,n-1}^\dagger + c_{-k,n+1}^\dagger) e^{i2\theta_3} \rbrace + M_A a_{k,n},\label{eq:subeq1} \\
        E_k & a_{-k,n}^\dagger= -(J + F/2) \lbrace( b_{-k,n}^\dagger e^{i\eta_1 k} + c_{-k,n}^\dagger e^{-i\eta_2 k} ) \nonumber \\
        &  + ( b_{-k,n+1}^\dagger + c_{-k,n+1}^\dagger) + ( b_{-k,n-1}^\dagger + c_{-k,n-1}^\dagger)\rbrace -\nonumber \\
        &(F/2)\lbrace ( b_{-k,n} e^{i\eta_1 k} + c_{-k,n} e^{-i\eta_2 k}) e^{-i2\theta_1} \nonumber \\
        & + ( b_{-k,n+1} + c_{-k,n-1}) e^{-i2\theta_2} \nonumber\\
        &+ ( b_{-k,n-1} + c_{-k,n+1}) e^{-i2\theta_3}\rbrace + M_A a_{-k,n}^\dagger, \label{eq:subeq2}\\
        E_k &b_{k,n} = -(J + F/2)(a_{k,n} e^{-i\eta_1k} + a_{k,n-1}, a_{k, n+1})-\nonumber\\
        & (F/2)(a_{-k,n}^\dagger e^{i2\theta_1} e^{-i\eta_1k} + a_{-k, n-1}^\dagger e^{i2\theta_2} + a_{-k, n+1}^\dagger e^{i2\theta_3})\nonumber\\
        &-(D/i)\lbrace b_{k, n+2} + b_{k, n-1}(1 + e^{i(-1)^n k})\rbrace \nonumber\\
        &+(D/i)\lbrace b_{k, n-2} + b_{k, n+1}(1 + e^{i(-1)^n k})\rbrace +  M_B b_{k,n},\label{eq:subeq3}\\
        E_k &b_{-k,n}^\dagger = -(J + F/2)(a_{-k,n}^\dagger e^{-i\eta_1k} + a_{-k,n-1}^\dagger, a_{-k, n+1}^\dagger)-\nonumber\\
        & (F/2)(a_{k,n} e^{-i2\theta_1} e^{-i\eta_1k} + a_{k, n-1} e^{-i2\theta_2} + a_{k, n+1} e^{-i2\theta_3})\nonumber\\
        &+(D/i)\lbrace b_{-k, n+2}^\dagger + b_{-k, n-1}^\dagger(1 + e^{i(-1)^n k})\rbrace -\nonumber\\
        &(D/i)\lbrace b_{-k, n-2}^\dagger + b_{-k, n+1}^\dagger(1 + e^{i(-1)^n k})\rbrace +  M_B b_{-k,n}^\dagger,\label{eq:subeq4}\\
        E_k &c_{k,n} = -(J + F/2)(a_{k,n} e^{i\eta_2k} + a_{k,n-1} + a_{k, n+1})-\nonumber\\
        & (F/2)(a_{-k,n}^\dagger e^{i2\theta_1} e^{i\eta_2k} + a_{-k, n+1}^\dagger e^{i2\theta_2} + a_{-k, n-1}^\dagger e^{i2\theta_3})\nonumber\\
        &-(D/i)\lbrace c_{k, n-2} + c_{k, n+1}(1 + e^{i(-1)^n k})\rbrace \nonumber\\
        &+(D/i)\lbrace c_{k, n+2} + c_{k, n-1}(1 + e^{i(-1)^n k})\rbrace +  M_C c_{k,n},\label{eq:subeq5}\\
        E_k &c_{-k,n}^\dagger = -(J + F/2)(a_{-k,n}^\dagger e^{i\eta_2k} + a_{-k,n-1}^\dagger + a_{-k, n+1}^\dagger)\nonumber\\
        & -(F/2)(a_{k,n} e^{-i2\theta_1} e^{i\eta_2k} + a_{k, n+1} e^{-i2\theta_2} + a_{k, n-1} e^{-i2\theta_3})\nonumber\\
        &\lbrace c_{-k, n+2}^\dagger + c_{-k, n-1}^\dagger(1 + e^{i(-1)^n k})\rbrace +  M_C c_{-k,n}^\dagger,\label{eq:subeq6}
    \end{align}\label{edge_eq}
    \endgroup
\end{subequations}
where $a_{n,k}$, $b_{n,k}$, $c_{n,k}$ are the bosonic operators corresponding to $\text{A}$, $\text{B}$, $\text{C}$ sublattice, respectively and the wave vector $k$ is scaled with $\sqrt{3}a_0 k_y$ ($a_0$: lattice constant). Here, $n$ is the site index which can have any value between 1 to $N$, where $N$ is the total number of lattice sites taken along $x$-direction. Further, $\eta_1 = (1 + (-1)^n)/2$ and $\eta_2 = (1 - (-1)^n)/2$ in Eq.~\ref{edge_eq}.


\bibliographystyle{apsrev4-2}
\bibliography{main}

\begin{thebibliography}{64}%
\makeatletter
\providecommand \@ifxundefined [1]{%
 \@ifx{#1\undefined}
}%
\providecommand \@ifnum [1]{%
 \ifnum #1\expandafter \@firstoftwo
 \else \expandafter \@secondoftwo
 \fi
}%
\providecommand \@ifx [1]{%
 \ifx #1\expandafter \@firstoftwo
 \else \expandafter \@secondoftwo
 \fi
}%
\providecommand \natexlab [1]{#1}%
\providecommand \enquote  [1]{``#1''}%
\providecommand \bibnamefont  [1]{#1}%
\providecommand \bibfnamefont [1]{#1}%
\providecommand \citenamefont [1]{#1}%
\providecommand \href@noop [0]{\@secondoftwo}%
\providecommand \href [0]{\begingroup \@sanitize@url \@href}%
\providecommand \@href[1]{\@@startlink{#1}\@@href}%
\providecommand \@@href[1]{\endgroup#1\@@endlink}%
\providecommand \@sanitize@url [0]{\catcode `\\12\catcode `\$12\catcode
  `\&12\catcode `\#12\catcode `\^12\catcode `\_12\catcode `\%12\relax}%
\providecommand \@@startlink[1]{}%
\providecommand \@@endlink[0]{}%
\providecommand \url  [0]{\begingroup\@sanitize@url \@url }%
\providecommand \@url [1]{\endgroup\@href {#1}{\urlprefix }}%
\providecommand \urlprefix  [0]{URL }%
\providecommand \Eprint [0]{\href }%
\providecommand \doibase [0]{https://doi.org/}%
\providecommand \selectlanguage [0]{\@gobble}%
\providecommand \bibinfo  [0]{\@secondoftwo}%
\providecommand \bibfield  [0]{\@secondoftwo}%
\providecommand \translation [1]{[#1]}%
\providecommand \BibitemOpen [0]{}%
\providecommand \bibitemStop [0]{}%
\providecommand \bibitemNoStop [0]{.\EOS\space}%
\providecommand \EOS [0]{\spacefactor3000\relax}%
\providecommand \BibitemShut  [1]{\csname bibitem#1\endcsname}%
\let\auto@bib@innerbib\@empty
\bibitem [{\citenamefont {Ye}\ and\ \citenamefont
  {Zou}(2024)}]{PhysRevX.14.021053}%
  \BibitemOpen
  \bibfield  {author} {\bibinfo {author} {\bibfnamefont {W.}~\bibnamefont
  {Ye}}\ and\ \bibinfo {author} {\bibfnamefont {L.}~\bibnamefont {Zou}},\
  }\href {https://doi.org/10.1103/PhysRevX.14.021053} {\bibfield  {journal}
  {\bibinfo  {journal} {Phys. Rev. X}\ }\textbf {\bibinfo {volume} {14}},\
  \bibinfo {pages} {021053} (\bibinfo {year} {2024})}\BibitemShut {NoStop}%
\bibitem [{\citenamefont {Dohi}\ \emph {et~al.}(2022)\citenamefont {Dohi},
  \citenamefont {Reeve},\ and\ \citenamefont {Kl{\"a}ui}}]{skyrmion3_2022}%
  \BibitemOpen
  \bibfield  {author} {\bibinfo {author} {\bibfnamefont {T.}~\bibnamefont
  {Dohi}}, \bibinfo {author} {\bibfnamefont {R.~M.}\ \bibnamefont {Reeve}},\
  and\ \bibinfo {author} {\bibfnamefont {M.}~\bibnamefont {Kl{\"a}ui}},\ }\href
  {https://doi.org/10.1146/annurev-conmatphys-031620-110344} {\bibfield
  {journal} {\bibinfo  {journal} {Annu. Rev. Condens. Matter Phys.}\ }\textbf
  {\bibinfo {volume} {13}},\ \bibinfo {pages} {73} (\bibinfo {year}
  {2022})}\BibitemShut {NoStop}%
\bibitem [{\citenamefont {Tang}\ \emph {et~al.}(2021)\citenamefont {Tang},
  \citenamefont {Wu}, \citenamefont {Wang}, \citenamefont {Kong}, \citenamefont
  {Lv}, \citenamefont {Wei}, \citenamefont {Zang}, \citenamefont {Tian},\ and\
  \citenamefont {Du}}]{skyrmion4_2021}%
  \BibitemOpen
  \bibfield  {author} {\bibinfo {author} {\bibfnamefont {J.}~\bibnamefont
  {Tang}}, \bibinfo {author} {\bibfnamefont {Y.}~\bibnamefont {Wu}}, \bibinfo
  {author} {\bibfnamefont {W.}~\bibnamefont {Wang}}, \bibinfo {author}
  {\bibfnamefont {L.}~\bibnamefont {Kong}}, \bibinfo {author} {\bibfnamefont
  {B.}~\bibnamefont {Lv}}, \bibinfo {author} {\bibfnamefont {W.}~\bibnamefont
  {Wei}}, \bibinfo {author} {\bibfnamefont {J.}~\bibnamefont {Zang}}, \bibinfo
  {author} {\bibfnamefont {M.}~\bibnamefont {Tian}},\ and\ \bibinfo {author}
  {\bibfnamefont {H.}~\bibnamefont {Du}},\ }\href
  {https://doi.org/10.1038/s41565-021-00954-9} {\bibfield  {journal} {\bibinfo
  {journal} {Nat. Nanotechnol.}\ }\textbf {\bibinfo {volume} {16}},\ \bibinfo
  {pages} {1086} (\bibinfo {year} {2021})}\BibitemShut {NoStop}%
\bibitem [{\citenamefont {Li}\ \emph {et~al.}(2023{\natexlab{a}})\citenamefont
  {Li}, \citenamefont {Wang},\ and\ \citenamefont {Rasing}}]{skyrmion5_2023}%
  \BibitemOpen
  \bibfield  {author} {\bibinfo {author} {\bibfnamefont {S.}~\bibnamefont
  {Li}}, \bibinfo {author} {\bibfnamefont {X.}~\bibnamefont {Wang}},\ and\
  \bibinfo {author} {\bibfnamefont {T.}~\bibnamefont {Rasing}},\ }\href
  {https://doi.org/10.1002/idm2.12072} {\bibfield  {journal} {\bibinfo
  {journal} {Interdisciplinary Materials}\ }\textbf {\bibinfo {volume} {2}},\
  \bibinfo {pages} {260} (\bibinfo {year} {2023}{\natexlab{a}})}\BibitemShut
  {NoStop}%
\bibitem [{\citenamefont {Moulsdale}\ \emph {et~al.}(2019)\citenamefont
  {Moulsdale}, \citenamefont {Pantale{\'o}n}, \citenamefont {Carrillo-Bastos},\
  and\ \citenamefont {Xian}}]{moulsdale2019unconventional}%
  \BibitemOpen
  \bibfield  {author} {\bibinfo {author} {\bibfnamefont {C.}~\bibnamefont
  {Moulsdale}}, \bibinfo {author} {\bibfnamefont {P.~A.}\ \bibnamefont
  {Pantale{\'o}n}}, \bibinfo {author} {\bibfnamefont {R.}~\bibnamefont
  {Carrillo-Bastos}},\ and\ \bibinfo {author} {\bibfnamefont {Y.}~\bibnamefont
  {Xian}},\ }\href {https://doi.org/10.1103/PhysRevB.99.214424} {\bibfield
  {journal} {\bibinfo  {journal} {Phys. Rev. B}\ }\textbf {\bibinfo {volume}
  {99}},\ \bibinfo {pages} {214424} (\bibinfo {year} {2019})}\BibitemShut
  {NoStop}%
\bibitem [{\citenamefont {Zhuo}\ \emph {et~al.}(2021)\citenamefont {Zhuo},
  \citenamefont {Li}, \citenamefont {Manchon} \emph
  {et~al.}}]{zhuo2021topological}%
  \BibitemOpen
  \bibfield  {author} {\bibinfo {author} {\bibfnamefont {F.}~\bibnamefont
  {Zhuo}}, \bibinfo {author} {\bibfnamefont {H.}~\bibnamefont {Li}}, \bibinfo
  {author} {\bibfnamefont {A.}~\bibnamefont {Manchon}}, \emph {et~al.},\ }\href
  {https://doi.org/10.1103/PhysRevB.104.144422} {\bibfield  {journal} {\bibinfo
   {journal} {Phys. Rev. B}\ }\textbf {\bibinfo {volume} {104}},\ \bibinfo
  {pages} {144422} (\bibinfo {year} {2021})}\BibitemShut {NoStop}%
\bibitem [{\citenamefont {Li}\ and\ \citenamefont
  {Cheng}(2021)}]{li2021magnonic}%
  \BibitemOpen
  \bibfield  {author} {\bibinfo {author} {\bibfnamefont {Y.-H.}\ \bibnamefont
  {Li}}\ and\ \bibinfo {author} {\bibfnamefont {R.}~\bibnamefont {Cheng}},\
  }\href {https://doi.org/10.1103/PhysRevB.103.014407} {\bibfield  {journal}
  {\bibinfo  {journal} {Phys. Rev. B}\ }\textbf {\bibinfo {volume} {103}},\
  \bibinfo {pages} {014407} (\bibinfo {year} {2021})}\BibitemShut {NoStop}%
\bibitem [{\citenamefont {Czajka}\ \emph {et~al.}(2023)\citenamefont {Czajka},
  \citenamefont {Gao}, \citenamefont {Hirschberger}, \citenamefont
  {Lampen-Kelley}, \citenamefont {Banerjee}, \citenamefont {Quirk},
  \citenamefont {Mandrus}, \citenamefont {Nagler},\ and\ \citenamefont
  {Ong}}]{czajka2023planar}%
  \BibitemOpen
  \bibfield  {author} {\bibinfo {author} {\bibfnamefont {P.}~\bibnamefont
  {Czajka}}, \bibinfo {author} {\bibfnamefont {T.}~\bibnamefont {Gao}},
  \bibinfo {author} {\bibfnamefont {M.}~\bibnamefont {Hirschberger}}, \bibinfo
  {author} {\bibfnamefont {P.}~\bibnamefont {Lampen-Kelley}}, \bibinfo {author}
  {\bibfnamefont {A.}~\bibnamefont {Banerjee}}, \bibinfo {author}
  {\bibfnamefont {N.}~\bibnamefont {Quirk}}, \bibinfo {author} {\bibfnamefont
  {D.~G.}\ \bibnamefont {Mandrus}}, \bibinfo {author} {\bibfnamefont {S.~E.}\
  \bibnamefont {Nagler}},\ and\ \bibinfo {author} {\bibfnamefont {N.~P.}\
  \bibnamefont {Ong}},\ }\href {https://doi.org/10.1038/s41563-022-01397-w}
  {\bibfield  {journal} {\bibinfo  {journal} {Nat. Mater.}\ }\textbf {\bibinfo
  {volume} {22}},\ \bibinfo {pages} {36} (\bibinfo {year} {2023})}\BibitemShut
  {NoStop}%
\bibitem [{\citenamefont {{\v{S}}mejkal}\ \emph {et~al.}(2018)\citenamefont
  {{\v{S}}mejkal}, \citenamefont {Mokrousov}, \citenamefont {Yan},\ and\
  \citenamefont {MacDonald}}]{spintronics1_2018}%
  \BibitemOpen
  \bibfield  {author} {\bibinfo {author} {\bibfnamefont {L.}~\bibnamefont
  {{\v{S}}mejkal}}, \bibinfo {author} {\bibfnamefont {Y.}~\bibnamefont
  {Mokrousov}}, \bibinfo {author} {\bibfnamefont {B.}~\bibnamefont {Yan}},\
  and\ \bibinfo {author} {\bibfnamefont {A.~H.}\ \bibnamefont {MacDonald}},\
  }\href {https://doi.org/10.1038/s41567-018-0064-5} {\bibfield  {journal}
  {\bibinfo  {journal} {Nat. phys.}\ }\textbf {\bibinfo {volume} {14}},\
  \bibinfo {pages} {242} (\bibinfo {year} {2018})}\BibitemShut {NoStop}%
\bibitem [{\citenamefont {Jungfleisch}(2022)}]{spintronics2_2022}%
  \BibitemOpen
  \bibfield  {author} {\bibinfo {author} {\bibfnamefont {M.~B.}\ \bibnamefont
  {Jungfleisch}},\ }\href {https://doi.org/10.1038/s41563-022-01416-w}
  {\bibfield  {journal} {\bibinfo  {journal} {Nat. Mater.}\ }\textbf {\bibinfo
  {volume} {21}},\ \bibinfo {pages} {1348} (\bibinfo {year}
  {2022})}\BibitemShut {NoStop}%
\bibitem [{\citenamefont {Han}\ \emph {et~al.}(2023)\citenamefont {Han},
  \citenamefont {Cheng}, \citenamefont {Liu}, \citenamefont {Ohno},\ and\
  \citenamefont {Fukami}}]{spintronics4_2023}%
  \BibitemOpen
  \bibfield  {author} {\bibinfo {author} {\bibfnamefont {J.}~\bibnamefont
  {Han}}, \bibinfo {author} {\bibfnamefont {R.}~\bibnamefont {Cheng}}, \bibinfo
  {author} {\bibfnamefont {L.}~\bibnamefont {Liu}}, \bibinfo {author}
  {\bibfnamefont {H.}~\bibnamefont {Ohno}},\ and\ \bibinfo {author}
  {\bibfnamefont {S.}~\bibnamefont {Fukami}},\ }\href
  {https://doi.org/10.1038/s41563-023-01492-6} {\bibfield  {journal} {\bibinfo
  {journal} {Nat. Mater.}\ }\textbf {\bibinfo {volume} {22}},\ \bibinfo {pages}
  {684} (\bibinfo {year} {2023})}\BibitemShut {NoStop}%
\bibitem [{\citenamefont {Hirohata}\ \emph {et~al.}(2020)\citenamefont
  {Hirohata}, \citenamefont {Yamada}, \citenamefont {Nakatani}, \citenamefont
  {Prejbeanu}, \citenamefont {Di{\'e}ny}, \citenamefont {Pirro},\ and\
  \citenamefont {Hillebrands}}]{spintronics5_2020}%
  \BibitemOpen
  \bibfield  {author} {\bibinfo {author} {\bibfnamefont {A.}~\bibnamefont
  {Hirohata}}, \bibinfo {author} {\bibfnamefont {K.}~\bibnamefont {Yamada}},
  \bibinfo {author} {\bibfnamefont {Y.}~\bibnamefont {Nakatani}}, \bibinfo
  {author} {\bibfnamefont {I.-L.}\ \bibnamefont {Prejbeanu}}, \bibinfo {author}
  {\bibfnamefont {B.}~\bibnamefont {Di{\'e}ny}}, \bibinfo {author}
  {\bibfnamefont {P.}~\bibnamefont {Pirro}},\ and\ \bibinfo {author}
  {\bibfnamefont {B.}~\bibnamefont {Hillebrands}},\ }\href
  {https://doi.org/10.1016/j.jmmm.2020.166711} {\bibfield  {journal} {\bibinfo
  {journal} {J. Magn. Magn. Mater.}\ }\textbf {\bibinfo {volume} {509}},\
  \bibinfo {pages} {166711} (\bibinfo {year} {2020})}\BibitemShut {NoStop}%
\bibitem [{\citenamefont {Lachance-Quirion}\ \emph {et~al.}(2019)\citenamefont
  {Lachance-Quirion}, \citenamefont {Tabuchi}, \citenamefont {Gloppe},
  \citenamefont {Usami},\ and\ \citenamefont {Nakamura}}]{spintronics6_2019}%
  \BibitemOpen
  \bibfield  {author} {\bibinfo {author} {\bibfnamefont {D.}~\bibnamefont
  {Lachance-Quirion}}, \bibinfo {author} {\bibfnamefont {Y.}~\bibnamefont
  {Tabuchi}}, \bibinfo {author} {\bibfnamefont {A.}~\bibnamefont {Gloppe}},
  \bibinfo {author} {\bibfnamefont {K.}~\bibnamefont {Usami}},\ and\ \bibinfo
  {author} {\bibfnamefont {Y.}~\bibnamefont {Nakamura}},\ }\href
  {https://doi.org/10.7567/1882-0786/ab248d} {\bibfield  {journal} {\bibinfo
  {journal} {Appl. Phys. Express}\ }\textbf {\bibinfo {volume} {12}},\ \bibinfo
  {pages} {070101} (\bibinfo {year} {2019})}\BibitemShut {NoStop}%
\bibitem [{\citenamefont {Kawano}\ and\ \citenamefont
  {Hotta}(2019{\natexlab{a}})}]{thermalHallantiferro}%
  \BibitemOpen
  \bibfield  {author} {\bibinfo {author} {\bibfnamefont {M.}~\bibnamefont
  {Kawano}}\ and\ \bibinfo {author} {\bibfnamefont {C.}~\bibnamefont {Hotta}},\
  }\href {https://doi.org/10.1103/PhysRevB.99.054422} {\bibfield  {journal}
  {\bibinfo  {journal} {Phys. Rev. B}\ }\textbf {\bibinfo {volume} {99}},\
  \bibinfo {pages} {054422} (\bibinfo {year} {2019}{\natexlab{a}})}\BibitemShut
  {NoStop}%
\bibitem [{\citenamefont {Sugii}\ \emph {et~al.}(2017)\citenamefont {Sugii},
  \citenamefont {Shimozawa}, \citenamefont {Watanabe}, \citenamefont {Suzuki},
  \citenamefont {Halim}, \citenamefont {Kimata}, \citenamefont {Matsumoto},
  \citenamefont {Nakatsuji},\ and\ \citenamefont
  {Yamashita}}]{sugii2017thermal}%
  \BibitemOpen
  \bibfield  {author} {\bibinfo {author} {\bibfnamefont {K.}~\bibnamefont
  {Sugii}}, \bibinfo {author} {\bibfnamefont {M.}~\bibnamefont {Shimozawa}},
  \bibinfo {author} {\bibfnamefont {D.}~\bibnamefont {Watanabe}}, \bibinfo
  {author} {\bibfnamefont {Y.}~\bibnamefont {Suzuki}}, \bibinfo {author}
  {\bibfnamefont {M.}~\bibnamefont {Halim}}, \bibinfo {author} {\bibfnamefont
  {M.}~\bibnamefont {Kimata}}, \bibinfo {author} {\bibfnamefont
  {Y.}~\bibnamefont {Matsumoto}}, \bibinfo {author} {\bibfnamefont
  {S.}~\bibnamefont {Nakatsuji}},\ and\ \bibinfo {author} {\bibfnamefont
  {M.}~\bibnamefont {Yamashita}},\ }\href
  {https://doi.org/10.1103/PhysRevLett.118.145902} {\bibfield  {journal}
  {\bibinfo  {journal} {Phys. Rev, Lett.}\ }\textbf {\bibinfo {volume} {118}},\
  \bibinfo {pages} {145902} (\bibinfo {year} {2017})}\BibitemShut {NoStop}%
\bibitem [{\citenamefont {Hirokane}\ \emph {et~al.}(2019)\citenamefont
  {Hirokane}, \citenamefont {Nii}, \citenamefont {Tomioka},\ and\ \citenamefont
  {Onose}}]{hirokane2019phononic}%
  \BibitemOpen
  \bibfield  {author} {\bibinfo {author} {\bibfnamefont {Y.}~\bibnamefont
  {Hirokane}}, \bibinfo {author} {\bibfnamefont {Y.}~\bibnamefont {Nii}},
  \bibinfo {author} {\bibfnamefont {Y.}~\bibnamefont {Tomioka}},\ and\ \bibinfo
  {author} {\bibfnamefont {Y.}~\bibnamefont {Onose}},\ }\href
  {https://doi.org/10.1103/PhysRevB.99.134419} {\bibfield  {journal} {\bibinfo
  {journal} {Phys. Rev. B}\ }\textbf {\bibinfo {volume} {99}},\ \bibinfo
  {pages} {134419} (\bibinfo {year} {2019})}\BibitemShut {NoStop}%
\bibitem [{\citenamefont {Li}\ \emph {et~al.}(2020)\citenamefont {Li},
  \citenamefont {Fauqu{\'e}}, \citenamefont {Zhu},\ and\ \citenamefont
  {Behnia}}]{li2020phonon}%
  \BibitemOpen
  \bibfield  {author} {\bibinfo {author} {\bibfnamefont {X.}~\bibnamefont
  {Li}}, \bibinfo {author} {\bibfnamefont {B.}~\bibnamefont {Fauqu{\'e}}},
  \bibinfo {author} {\bibfnamefont {Z.}~\bibnamefont {Zhu}},\ and\ \bibinfo
  {author} {\bibfnamefont {K.}~\bibnamefont {Behnia}},\ }\href
  {https://doi.org/10.1103/PhysRevLett.124.105901} {\bibfield  {journal}
  {\bibinfo  {journal} {Phys. Rev. Lett.}\ }\textbf {\bibinfo {volume} {124}},\
  \bibinfo {pages} {105901} (\bibinfo {year} {2020})}\BibitemShut {NoStop}%
\bibitem [{\citenamefont {Katsura}\ \emph {et~al.}(2010)\citenamefont
  {Katsura}, \citenamefont {Nagaosa},\ and\ \citenamefont
  {Lee}}]{thermalhall2010}%
  \BibitemOpen
  \bibfield  {author} {\bibinfo {author} {\bibfnamefont {H.}~\bibnamefont
  {Katsura}}, \bibinfo {author} {\bibfnamefont {N.}~\bibnamefont {Nagaosa}},\
  and\ \bibinfo {author} {\bibfnamefont {P.~A.}\ \bibnamefont {Lee}},\ }\href
  {https://doi.org/10.1103/PhysRevLett.104.066403} {\bibfield  {journal}
  {\bibinfo  {journal} {Phys. Rev. Lett.}\ }\textbf {\bibinfo {volume} {104}},\
  \bibinfo {pages} {066403} (\bibinfo {year} {2010})}\BibitemShut {NoStop}%
\bibitem [{\citenamefont {Jin}\ \emph {et~al.}(2015)\citenamefont {Jin},
  \citenamefont {Boona}, \citenamefont {Yang}, \citenamefont {Myers},\ and\
  \citenamefont {Heremans}}]{Seebeck}%
  \BibitemOpen
  \bibfield  {author} {\bibinfo {author} {\bibfnamefont {H.}~\bibnamefont
  {Jin}}, \bibinfo {author} {\bibfnamefont {S.~R.}\ \bibnamefont {Boona}},
  \bibinfo {author} {\bibfnamefont {Z.}~\bibnamefont {Yang}}, \bibinfo {author}
  {\bibfnamefont {R.~C.}\ \bibnamefont {Myers}},\ and\ \bibinfo {author}
  {\bibfnamefont {J.~P.}\ \bibnamefont {Heremans}},\ }\href
  {https://doi.org/10.1103/PhysRevB.92.054436} {\bibfield  {journal} {\bibinfo
  {journal} {Phys. Rev. B}\ }\textbf {\bibinfo {volume} {92}},\ \bibinfo
  {pages} {054436} (\bibinfo {year} {2015})}\BibitemShut {NoStop}%
\bibitem [{\citenamefont {Cheng}\ \emph {et~al.}(2016)\citenamefont {Cheng},
  \citenamefont {Okamoto},\ and\ \citenamefont {Xiao}}]{NernstEffect1}%
  \BibitemOpen
  \bibfield  {author} {\bibinfo {author} {\bibfnamefont {R.}~\bibnamefont
  {Cheng}}, \bibinfo {author} {\bibfnamefont {S.}~\bibnamefont {Okamoto}},\
  and\ \bibinfo {author} {\bibfnamefont {D.}~\bibnamefont {Xiao}},\ }\href
  {https://doi.org/10.1103/PhysRevLett.117.217202} {\bibfield  {journal}
  {\bibinfo  {journal} {Phys. Rev. Lett.}\ }\textbf {\bibinfo {volume} {117}},\
  \bibinfo {pages} {217202} (\bibinfo {year} {2016})}\BibitemShut {NoStop}%
\bibitem [{\citenamefont {Zyuzin}\ and\ \citenamefont
  {Kovalev}(2016)}]{NernstEffect2}%
  \BibitemOpen
  \bibfield  {author} {\bibinfo {author} {\bibfnamefont {V.~A.}\ \bibnamefont
  {Zyuzin}}\ and\ \bibinfo {author} {\bibfnamefont {A.~A.}\ \bibnamefont
  {Kovalev}},\ }\href {https://doi.org/10.1103/PhysRevLett.117.217203}
  {\bibfield  {journal} {\bibinfo  {journal} {Phys. Rev. Lett.}\ }\textbf
  {\bibinfo {volume} {117}},\ \bibinfo {pages} {217203} (\bibinfo {year}
  {2016})}\BibitemShut {NoStop}%
\bibitem [{\citenamefont {Shiomi}\ \emph {et~al.}(2017)\citenamefont {Shiomi},
  \citenamefont {Takashima},\ and\ \citenamefont {Saitoh}}]{NernstEffect3}%
  \BibitemOpen
  \bibfield  {author} {\bibinfo {author} {\bibfnamefont {Y.}~\bibnamefont
  {Shiomi}}, \bibinfo {author} {\bibfnamefont {R.}~\bibnamefont {Takashima}},\
  and\ \bibinfo {author} {\bibfnamefont {E.}~\bibnamefont {Saitoh}},\ }\href
  {https://doi.org/10.1103/PhysRevB.96.134425} {\bibfield  {journal} {\bibinfo
  {journal} {Phys. Rev. B}\ }\textbf {\bibinfo {volume} {96}},\ \bibinfo
  {pages} {134425} (\bibinfo {year} {2017})}\BibitemShut {NoStop}%
\bibitem [{\citenamefont {Zhang}\ \emph {et~al.}(2010)\citenamefont {Zhang},
  \citenamefont {Ren}, \citenamefont {Wang},\ and\ \citenamefont
  {Li}}]{PhononHall}%
  \BibitemOpen
  \bibfield  {author} {\bibinfo {author} {\bibfnamefont {L.}~\bibnamefont
  {Zhang}}, \bibinfo {author} {\bibfnamefont {J.}~\bibnamefont {Ren}}, \bibinfo
  {author} {\bibfnamefont {J.-S.}\ \bibnamefont {Wang}},\ and\ \bibinfo
  {author} {\bibfnamefont {B.}~\bibnamefont {Li}},\ }\href
  {https://doi.org/10.1103/PhysRevLett.105.225901} {\bibfield  {journal}
  {\bibinfo  {journal} {Phys. Rev. Lett.}\ }\textbf {\bibinfo {volume} {105}},\
  \bibinfo {pages} {225901} (\bibinfo {year} {2010})}\BibitemShut {NoStop}%
\bibitem [{\citenamefont {Haldane}\ and\ \citenamefont
  {Raghu}(2008)}]{PhotonHall}%
  \BibitemOpen
  \bibfield  {author} {\bibinfo {author} {\bibfnamefont {F.~D.~M.}\
  \bibnamefont {Haldane}}\ and\ \bibinfo {author} {\bibfnamefont
  {S.}~\bibnamefont {Raghu}},\ }\href
  {https://doi.org/10.1103/PhysRevLett.100.013904} {\bibfield  {journal}
  {\bibinfo  {journal} {Phys. Rev. Lett.}\ }\textbf {\bibinfo {volume} {100}},\
  \bibinfo {pages} {013904} (\bibinfo {year} {2008})}\BibitemShut {NoStop}%
\bibitem [{\citenamefont {Li}\ \emph {et~al.}(2022)\citenamefont {Li},
  \citenamefont {Ma}, \citenamefont {Chen}, \citenamefont {Xie},\ and\
  \citenamefont {Ma}}]{magnon-skyrmion}%
  \BibitemOpen
  \bibfield  {author} {\bibinfo {author} {\bibfnamefont {Z.}~\bibnamefont
  {Li}}, \bibinfo {author} {\bibfnamefont {M.}~\bibnamefont {Ma}}, \bibinfo
  {author} {\bibfnamefont {Z.}~\bibnamefont {Chen}}, \bibinfo {author}
  {\bibfnamefont {K.}~\bibnamefont {Xie}},\ and\ \bibinfo {author}
  {\bibfnamefont {F.}~\bibnamefont {Ma}},\ }\href
  {https://doi.org/10.1063/5.0121314} {\bibfield  {journal} {\bibinfo
  {journal} {J. Appl. Phys.}\ }\textbf {\bibinfo {volume} {132}} (\bibinfo
  {year} {2022})}\BibitemShut {NoStop}%
\bibitem [{\citenamefont {Yuan}\ \emph {et~al.}(2022)\citenamefont {Yuan},
  \citenamefont {Cao}, \citenamefont {Kamra}, \citenamefont {Duine},\ and\
  \citenamefont {Yan}}]{magnon1}%
  \BibitemOpen
  \bibfield  {author} {\bibinfo {author} {\bibfnamefont {H.}~\bibnamefont
  {Yuan}}, \bibinfo {author} {\bibfnamefont {Y.}~\bibnamefont {Cao}}, \bibinfo
  {author} {\bibfnamefont {A.}~\bibnamefont {Kamra}}, \bibinfo {author}
  {\bibfnamefont {R.~A.}\ \bibnamefont {Duine}},\ and\ \bibinfo {author}
  {\bibfnamefont {P.}~\bibnamefont {Yan}},\ }\href
  {https://doi.org/10.1016/j.physrep.2022.03.002} {\bibfield  {journal}
  {\bibinfo  {journal} {Phys. Rep.}\ }\textbf {\bibinfo {volume} {965}},\
  \bibinfo {pages} {1} (\bibinfo {year} {2022})}\BibitemShut {NoStop}%
\bibitem [{\citenamefont {Yang}\ \emph {et~al.}(2023)\citenamefont {Yang},
  \citenamefont {Zhao}, \citenamefont {Wu}, \citenamefont {Chu}, \citenamefont
  {Xu}, \citenamefont {Li}, \citenamefont {{\AA}kerman},\ and\ \citenamefont
  {Zhou}}]{skyrmion1}%
  \BibitemOpen
  \bibfield  {author} {\bibinfo {author} {\bibfnamefont {S.}~\bibnamefont
  {Yang}}, \bibinfo {author} {\bibfnamefont {Y.}~\bibnamefont {Zhao}}, \bibinfo
  {author} {\bibfnamefont {K.}~\bibnamefont {Wu}}, \bibinfo {author}
  {\bibfnamefont {Z.}~\bibnamefont {Chu}}, \bibinfo {author} {\bibfnamefont
  {X.}~\bibnamefont {Xu}}, \bibinfo {author} {\bibfnamefont {X.}~\bibnamefont
  {Li}}, \bibinfo {author} {\bibfnamefont {J.}~\bibnamefont {{\AA}kerman}},\
  and\ \bibinfo {author} {\bibfnamefont {Y.}~\bibnamefont {Zhou}},\ }\href
  {https://doi.org/10.1038/s41467-023-39007-1} {\bibfield  {journal} {\bibinfo
  {journal} {Nat. Commun.}\ }\textbf {\bibinfo {volume} {14}},\ \bibinfo
  {pages} {3406} (\bibinfo {year} {2023})}\BibitemShut {NoStop}%
\bibitem [{\citenamefont {Li}\ \emph {et~al.}(2023{\natexlab{b}})\citenamefont
  {Li}, \citenamefont {Wang},\ and\ \citenamefont {Rasing}}]{skyrmion2}%
  \BibitemOpen
  \bibfield  {author} {\bibinfo {author} {\bibfnamefont {S.}~\bibnamefont
  {Li}}, \bibinfo {author} {\bibfnamefont {X.}~\bibnamefont {Wang}},\ and\
  \bibinfo {author} {\bibfnamefont {T.}~\bibnamefont {Rasing}},\ }\href
  {https://doi.org/10.1002/idm2.12072} {\bibfield  {journal} {\bibinfo
  {journal} {Interdisciplinary Materials}\ }\textbf {\bibinfo {volume} {2}},\
  \bibinfo {pages} {260} (\bibinfo {year} {2023}{\natexlab{b}})}\BibitemShut
  {NoStop}%
\bibitem [{\citenamefont {Shekhtman}\ \emph {et~al.}(1992)\citenamefont
  {Shekhtman}, \citenamefont {Entin-Wohlman},\ and\ \citenamefont
  {Aharony}}]{PhysRevLett.69.836}%
  \BibitemOpen
  \bibfield  {author} {\bibinfo {author} {\bibfnamefont {L.}~\bibnamefont
  {Shekhtman}}, \bibinfo {author} {\bibfnamefont {O.}~\bibnamefont
  {Entin-Wohlman}},\ and\ \bibinfo {author} {\bibfnamefont {A.}~\bibnamefont
  {Aharony}},\ }\href {https://doi.org/10.1103/PhysRevLett.69.836} {\bibfield
  {journal} {\bibinfo  {journal} {Phys. Rev. Lett.}\ }\textbf {\bibinfo
  {volume} {69}},\ \bibinfo {pages} {836} (\bibinfo {year} {1992})}\BibitemShut
  {NoStop}%
\bibitem [{\citenamefont {Jackeli}\ and\ \citenamefont
  {Khaliullin}(2009)}]{PhysRevLett.102.017205}%
  \BibitemOpen
  \bibfield  {author} {\bibinfo {author} {\bibfnamefont {G.}~\bibnamefont
  {Jackeli}}\ and\ \bibinfo {author} {\bibfnamefont {G.}~\bibnamefont
  {Khaliullin}},\ }\href {https://doi.org/10.1103/PhysRevLett.102.017205}
  {\bibfield  {journal} {\bibinfo  {journal} {Phys. Rev. Lett.}\ }\textbf
  {\bibinfo {volume} {102}},\ \bibinfo {pages} {017205} (\bibinfo {year}
  {2009})}\BibitemShut {NoStop}%
\bibitem [{\citenamefont {Debnath}\ and\ \citenamefont
  {Basu}(2023)}]{debnath2023multiple}%
  \BibitemOpen
  \bibfield  {author} {\bibinfo {author} {\bibfnamefont {S.}~\bibnamefont
  {Debnath}}\ and\ \bibinfo {author} {\bibfnamefont {S.}~\bibnamefont {Basu}},\
  }\href {https://arxiv.org/abs/2311.07201} {\bibfield  {journal} {\bibinfo
  {journal} {arXiv preprint arXiv:2311.07201}\ } (\bibinfo {year}
  {2023})}\BibitemShut {NoStop}%
\bibitem [{\citenamefont {Kawano}\ and\ \citenamefont
  {Hotta}(2019{\natexlab{b}})}]{square1}%
  \BibitemOpen
  \bibfield  {author} {\bibinfo {author} {\bibfnamefont {M.}~\bibnamefont
  {Kawano}}\ and\ \bibinfo {author} {\bibfnamefont {C.}~\bibnamefont {Hotta}},\
  }\href {https://doi.org/10.1103/PhysRevB.99.054422} {\bibfield  {journal}
  {\bibinfo  {journal} {Phys. Rev. B}\ }\textbf {\bibinfo {volume} {99}},\
  \bibinfo {pages} {054422} (\bibinfo {year} {2019}{\natexlab{b}})}\BibitemShut
  {NoStop}%
\bibitem [{\citenamefont {Kim}\ and\ \citenamefont {Kim}(2022)}]{honeycomb1}%
  \BibitemOpen
  \bibfield  {author} {\bibinfo {author} {\bibfnamefont {H.}~\bibnamefont
  {Kim}}\ and\ \bibinfo {author} {\bibfnamefont {S.~K.}\ \bibnamefont {Kim}},\
  }\href {https://doi.org/10.1103/PhysRevB.106.104430} {\bibfield  {journal}
  {\bibinfo  {journal} {Phys. Rev. B}\ }\textbf {\bibinfo {volume} {106}},\
  \bibinfo {pages} {104430} (\bibinfo {year} {2022})}\BibitemShut {NoStop}%
\bibitem [{\citenamefont {Laurell}\ and\ \citenamefont
  {Fiete}(2018)}]{PhysRevB.98.094419}%
  \BibitemOpen
  \bibfield  {author} {\bibinfo {author} {\bibfnamefont {P.}~\bibnamefont
  {Laurell}}\ and\ \bibinfo {author} {\bibfnamefont {G.~A.}\ \bibnamefont
  {Fiete}},\ }\href {https://doi.org/10.1103/PhysRevB.98.094419} {\bibfield
  {journal} {\bibinfo  {journal} {Phys. Rev. B}\ }\textbf {\bibinfo {volume}
  {98}},\ \bibinfo {pages} {094419} (\bibinfo {year} {2018})}\BibitemShut
  {NoStop}%
\bibitem [{\citenamefont {Akazawa}\ \emph {et~al.}(2020)\citenamefont
  {Akazawa}, \citenamefont {Shimozawa}, \citenamefont {Kittaka}, \citenamefont
  {Sakakibara}, \citenamefont {Okuma}, \citenamefont {Hiroi}, \citenamefont
  {Lee}, \citenamefont {Kawashima}, \citenamefont {Han},\ and\ \citenamefont
  {Yamashita}}]{PhysRevX.10.041059}%
  \BibitemOpen
  \bibfield  {author} {\bibinfo {author} {\bibfnamefont {M.}~\bibnamefont
  {Akazawa}}, \bibinfo {author} {\bibfnamefont {M.}~\bibnamefont {Shimozawa}},
  \bibinfo {author} {\bibfnamefont {S.}~\bibnamefont {Kittaka}}, \bibinfo
  {author} {\bibfnamefont {T.}~\bibnamefont {Sakakibara}}, \bibinfo {author}
  {\bibfnamefont {R.}~\bibnamefont {Okuma}}, \bibinfo {author} {\bibfnamefont
  {Z.}~\bibnamefont {Hiroi}}, \bibinfo {author} {\bibfnamefont {H.-Y.}\
  \bibnamefont {Lee}}, \bibinfo {author} {\bibfnamefont {N.}~\bibnamefont
  {Kawashima}}, \bibinfo {author} {\bibfnamefont {J.~H.}\ \bibnamefont {Han}},\
  and\ \bibinfo {author} {\bibfnamefont {M.}~\bibnamefont {Yamashita}},\ }\href
  {https://doi.org/10.1103/PhysRevX.10.041059} {\bibfield  {journal} {\bibinfo
  {journal} {Phys. Rev. X}\ }\textbf {\bibinfo {volume} {10}},\ \bibinfo
  {pages} {041059} (\bibinfo {year} {2020})}\BibitemShut {NoStop}%
\bibitem [{\citenamefont {Mook}\ \emph
  {et~al.}(2014{\natexlab{a}})\citenamefont {Mook}, \citenamefont {Henk},\ and\
  \citenamefont {Mertig}}]{PhysRevB.89.134409}%
  \BibitemOpen
  \bibfield  {author} {\bibinfo {author} {\bibfnamefont {A.}~\bibnamefont
  {Mook}}, \bibinfo {author} {\bibfnamefont {J.}~\bibnamefont {Henk}},\ and\
  \bibinfo {author} {\bibfnamefont {I.}~\bibnamefont {Mertig}},\ }\href
  {https://doi.org/10.1103/PhysRevB.89.134409} {\bibfield  {journal} {\bibinfo
  {journal} {Phys. Rev. B}\ }\textbf {\bibinfo {volume} {89}},\ \bibinfo
  {pages} {134409} (\bibinfo {year} {2014}{\natexlab{a}})}\BibitemShut
  {NoStop}%
\bibitem [{\citenamefont {Laurell}\ and\ \citenamefont
  {Fiete}(2017)}]{PhysRevLett.118.177201}%
  \BibitemOpen
  \bibfield  {author} {\bibinfo {author} {\bibfnamefont {P.}~\bibnamefont
  {Laurell}}\ and\ \bibinfo {author} {\bibfnamefont {G.~A.}\ \bibnamefont
  {Fiete}},\ }\href {https://doi.org/10.1103/PhysRevLett.118.177201} {\bibfield
   {journal} {\bibinfo  {journal} {Phys. Rev. Lett.}\ }\textbf {\bibinfo
  {volume} {118}},\ \bibinfo {pages} {177201} (\bibinfo {year}
  {2017})}\BibitemShut {NoStop}%
\bibitem [{\citenamefont {Wang}\ and\ \citenamefont
  {Ran}(2011)}]{PhysRevB.84.241103}%
  \BibitemOpen
  \bibfield  {author} {\bibinfo {author} {\bibfnamefont {F.}~\bibnamefont
  {Wang}}\ and\ \bibinfo {author} {\bibfnamefont {Y.}~\bibnamefont {Ran}},\
  }\href {https://doi.org/10.1103/PhysRevB.84.241103} {\bibfield  {journal}
  {\bibinfo  {journal} {Phys. Rev. B}\ }\textbf {\bibinfo {volume} {84}},\
  \bibinfo {pages} {241103} (\bibinfo {year} {2011})}\BibitemShut {NoStop}%
\bibitem [{\citenamefont {Mondal}\ and\ \citenamefont
  {Basu}(2023)}]{PhysRevB.107.035421}%
  \BibitemOpen
  \bibfield  {author} {\bibinfo {author} {\bibfnamefont {S.}~\bibnamefont
  {Mondal}}\ and\ \bibinfo {author} {\bibfnamefont {S.}~\bibnamefont {Basu}},\
  }\href {https://doi.org/10.1103/PhysRevB.107.035421} {\bibfield  {journal}
  {\bibinfo  {journal} {Phys. Rev. B}\ }\textbf {\bibinfo {volume} {107}},\
  \bibinfo {pages} {035421} (\bibinfo {year} {2023})}\BibitemShut {NoStop}%
\bibitem [{\citenamefont {Xiao}\ \emph {et~al.}(2011)\citenamefont {Xiao},
  \citenamefont {Zhu}, \citenamefont {Ran}, \citenamefont {Nagaosa},\ and\
  \citenamefont {Okamoto}}]{xiao2011interface}%
  \BibitemOpen
  \bibfield  {author} {\bibinfo {author} {\bibfnamefont {D.}~\bibnamefont
  {Xiao}}, \bibinfo {author} {\bibfnamefont {W.}~\bibnamefont {Zhu}}, \bibinfo
  {author} {\bibfnamefont {Y.}~\bibnamefont {Ran}}, \bibinfo {author}
  {\bibfnamefont {N.}~\bibnamefont {Nagaosa}},\ and\ \bibinfo {author}
  {\bibfnamefont {S.}~\bibnamefont {Okamoto}},\ }\href
  {https://doi.org/https://doi.org/10.1038/ncomms1602} {\bibfield  {journal}
  {\bibinfo  {journal} {Nat. commun.}\ }\textbf {\bibinfo {volume} {2}},\
  \bibinfo {pages} {596} (\bibinfo {year} {2011})}\BibitemShut {NoStop}%
\bibitem [{\citenamefont {Jagannathan}\ \emph {et~al.}(2006)\citenamefont
  {Jagannathan}, \citenamefont {Moessner},\ and\ \citenamefont
  {Wessel}}]{dice_magnet}%
  \BibitemOpen
  \bibfield  {author} {\bibinfo {author} {\bibfnamefont {A.}~\bibnamefont
  {Jagannathan}}, \bibinfo {author} {\bibfnamefont {R.}~\bibnamefont
  {Moessner}},\ and\ \bibinfo {author} {\bibfnamefont {S.}~\bibnamefont
  {Wessel}},\ }\href {https://doi.org/10.1103/PhysRevB.74.184410} {\bibfield
  {journal} {\bibinfo  {journal} {Phys. Rev. B}\ }\textbf {\bibinfo {volume}
  {74}},\ \bibinfo {pages} {184410} (\bibinfo {year} {2006})}\BibitemShut
  {NoStop}%
\bibitem [{\citenamefont {Jagannathan}\ \emph {et~al.}(2012)\citenamefont
  {Jagannathan}, \citenamefont {Dou\ifmmode~\mbox{\c{c}}\else \c{c}\fi{}ot},
  \citenamefont {Szallas},\ and\ \citenamefont {Wessel}}]{dice_magnet1}%
  \BibitemOpen
  \bibfield  {author} {\bibinfo {author} {\bibfnamefont {A.}~\bibnamefont
  {Jagannathan}}, \bibinfo {author} {\bibfnamefont {B.}~\bibnamefont
  {Dou\ifmmode~\mbox{\c{c}}\else \c{c}\fi{}ot}}, \bibinfo {author}
  {\bibfnamefont {A.}~\bibnamefont {Szallas}},\ and\ \bibinfo {author}
  {\bibfnamefont {S.}~\bibnamefont {Wessel}},\ }\href
  {https://doi.org/10.1103/PhysRevB.85.094434} {\bibfield  {journal} {\bibinfo
  {journal} {Phys. Rev. B}\ }\textbf {\bibinfo {volume} {85}},\ \bibinfo
  {pages} {094434} (\bibinfo {year} {2012})}\BibitemShut {NoStop}%
\bibitem [{\citenamefont {Mook}\ \emph {et~al.}(2019)\citenamefont {Mook},
  \citenamefont {Henk},\ and\ \citenamefont {Mertig}}]{PhysRevB.99.014427}%
  \BibitemOpen
  \bibfield  {author} {\bibinfo {author} {\bibfnamefont {A.}~\bibnamefont
  {Mook}}, \bibinfo {author} {\bibfnamefont {J.}~\bibnamefont {Henk}},\ and\
  \bibinfo {author} {\bibfnamefont {I.}~\bibnamefont {Mertig}},\ }\href
  {https://doi.org/10.1103/PhysRevB.99.014427} {\bibfield  {journal} {\bibinfo
  {journal} {Phys. Rev. B}\ }\textbf {\bibinfo {volume} {99}},\ \bibinfo
  {pages} {014427} (\bibinfo {year} {2019})}\BibitemShut {NoStop}%
\bibitem [{\citenamefont {Kim}\ \emph {et~al.}(2024)\citenamefont {Kim},
  \citenamefont {Saito}, \citenamefont {Yang}, \citenamefont {Ishizuka},
  \citenamefont {Coak}, \citenamefont {Lee}, \citenamefont {Sim}, \citenamefont
  {Oh}, \citenamefont {Nagaosa},\ and\ \citenamefont {Park}}]{kim2024thermal}%
  \BibitemOpen
  \bibfield  {author} {\bibinfo {author} {\bibfnamefont {H.-L.}\ \bibnamefont
  {Kim}}, \bibinfo {author} {\bibfnamefont {T.}~\bibnamefont {Saito}}, \bibinfo
  {author} {\bibfnamefont {H.}~\bibnamefont {Yang}}, \bibinfo {author}
  {\bibfnamefont {H.}~\bibnamefont {Ishizuka}}, \bibinfo {author}
  {\bibfnamefont {M.~J.}\ \bibnamefont {Coak}}, \bibinfo {author}
  {\bibfnamefont {J.~H.}\ \bibnamefont {Lee}}, \bibinfo {author} {\bibfnamefont
  {H.}~\bibnamefont {Sim}}, \bibinfo {author} {\bibfnamefont {Y.~S.}\
  \bibnamefont {Oh}}, \bibinfo {author} {\bibfnamefont {N.}~\bibnamefont
  {Nagaosa}},\ and\ \bibinfo {author} {\bibfnamefont {J.-G.}\ \bibnamefont
  {Park}},\ }\href {https://doi.org/https://doi.org/10.1038/s41467-023-44448-9}
  {\bibfield  {journal} {\bibinfo  {journal} {Nat. Commun.}\ }\textbf {\bibinfo
  {volume} {15}},\ \bibinfo {pages} {243} (\bibinfo {year} {2024})}\BibitemShut
  {NoStop}%
\bibitem [{\citenamefont {McClarty}(2022)}]{mcclarty2022topological}%
  \BibitemOpen
  \bibfield  {author} {\bibinfo {author} {\bibfnamefont {P.~A.}\ \bibnamefont
  {McClarty}},\ }\href
  {https://doi.org/10.1146/annurev-conmatphys-031620-104715} {\bibfield
  {journal} {\bibinfo  {journal} {Annu. Rev. Condens. Matter Phys.}\ }\textbf
  {\bibinfo {volume} {13}},\ \bibinfo {pages} {171} (\bibinfo {year}
  {2022})}\BibitemShut {NoStop}%
\bibitem [{\citenamefont {Mook}\ \emph {et~al.}(2023)\citenamefont {Mook},
  \citenamefont {Hoyer}, \citenamefont {Klinovaja},\ and\ \citenamefont
  {Loss}}]{magnonboundpairs2023}%
  \BibitemOpen
  \bibfield  {author} {\bibinfo {author} {\bibfnamefont {A.}~\bibnamefont
  {Mook}}, \bibinfo {author} {\bibfnamefont {R.}~\bibnamefont {Hoyer}},
  \bibinfo {author} {\bibfnamefont {J.}~\bibnamefont {Klinovaja}},\ and\
  \bibinfo {author} {\bibfnamefont {D.}~\bibnamefont {Loss}},\ }\href
  {https://doi.org/10.1103/PhysRevB.107.064429} {\bibfield  {journal} {\bibinfo
   {journal} {Phys. Rev. B}\ }\textbf {\bibinfo {volume} {107}},\ \bibinfo
  {pages} {064429} (\bibinfo {year} {2023})}\BibitemShut {NoStop}%
\bibitem [{\citenamefont {Singh}\ and\ \citenamefont
  {Sharma}(2023)}]{PhysRevB.107.245150}%
  \BibitemOpen
  \bibfield  {author} {\bibinfo {author} {\bibfnamefont {A.}~\bibnamefont
  {Singh}}\ and\ \bibinfo {author} {\bibfnamefont {G.}~\bibnamefont {Sharma}},\
  }\href {https://doi.org/10.1103/PhysRevB.107.245150} {\bibfield  {journal}
  {\bibinfo  {journal} {Phys. Rev. B}\ }\textbf {\bibinfo {volume} {107}},\
  \bibinfo {pages} {245150} (\bibinfo {year} {2023})}\BibitemShut {NoStop}%
\bibitem [{\citenamefont {Islam}\ \emph {et~al.}(2024)\citenamefont {Islam},
  \citenamefont {Bhattacharyya},\ and\ \citenamefont
  {Basu}}]{PhysRevB.110.045426}%
  \BibitemOpen
  \bibfield  {author} {\bibinfo {author} {\bibfnamefont {M.}~\bibnamefont
  {Islam}}, \bibinfo {author} {\bibfnamefont {K.}~\bibnamefont
  {Bhattacharyya}},\ and\ \bibinfo {author} {\bibfnamefont {S.}~\bibnamefont
  {Basu}},\ }\href {https://doi.org/10.1103/PhysRevB.110.045426} {\bibfield
  {journal} {\bibinfo  {journal} {Phys. Rev. B}\ }\textbf {\bibinfo {volume}
  {110}},\ \bibinfo {pages} {045426} (\bibinfo {year} {2024})}\BibitemShut
  {NoStop}%
\bibitem [{\citenamefont {Holstein}\ and\ \citenamefont
  {Primakoff}(1940)}]{holstein1940field}%
  \BibitemOpen
  \bibfield  {author} {\bibinfo {author} {\bibfnamefont {T.}~\bibnamefont
  {Holstein}}\ and\ \bibinfo {author} {\bibfnamefont {H.}~\bibnamefont
  {Primakoff}},\ }\href {https://doi.org/10.1103/PhysRev.58.1098} {\bibfield
  {journal} {\bibinfo  {journal} {Phys. Rev.}\ }\textbf {\bibinfo {volume}
  {58}},\ \bibinfo {pages} {1098} (\bibinfo {year} {1940})}\BibitemShut
  {NoStop}%
\bibitem [{\citenamefont {Wang}\ \emph {et~al.}(2020)\citenamefont {Wang},
  \citenamefont {Brataas},\ and\ \citenamefont
  {Troncoso}}]{PhysRevLett.125.217202}%
  \BibitemOpen
  \bibfield  {author} {\bibinfo {author} {\bibfnamefont {X.~S.}\ \bibnamefont
  {Wang}}, \bibinfo {author} {\bibfnamefont {A.}~\bibnamefont {Brataas}},\ and\
  \bibinfo {author} {\bibfnamefont {R.~E.}\ \bibnamefont {Troncoso}},\ }\href
  {https://doi.org/10.1103/PhysRevLett.125.217202} {\bibfield  {journal}
  {\bibinfo  {journal} {Phys. Rev. Lett.}\ }\textbf {\bibinfo {volume} {125}},\
  \bibinfo {pages} {217202} (\bibinfo {year} {2020})}\BibitemShut {NoStop}%
\bibitem [{\citenamefont {Yu}\ \emph {et~al.}(2019)\citenamefont {Yu},
  \citenamefont {Deng}, \citenamefont {Cao}, \citenamefont {McIntyre},
  \citenamefont {Li}, \citenamefont {Yuan}, \citenamefont {Feng}, \citenamefont
  {Ge}, \citenamefont {Zhang},\ and\ \citenamefont {Cao}}]{yu2019tuning}%
  \BibitemOpen
  \bibfield  {author} {\bibinfo {author} {\bibfnamefont {Y.}~\bibnamefont
  {Yu}}, \bibinfo {author} {\bibfnamefont {G.}~\bibnamefont {Deng}}, \bibinfo
  {author} {\bibfnamefont {Y.}~\bibnamefont {Cao}}, \bibinfo {author}
  {\bibfnamefont {G.~J.}\ \bibnamefont {McIntyre}}, \bibinfo {author}
  {\bibfnamefont {R.}~\bibnamefont {Li}}, \bibinfo {author} {\bibfnamefont
  {N.}~\bibnamefont {Yuan}}, \bibinfo {author} {\bibfnamefont {Z.}~\bibnamefont
  {Feng}}, \bibinfo {author} {\bibfnamefont {J.-Y.}\ \bibnamefont {Ge}},
  \bibinfo {author} {\bibfnamefont {J.}~\bibnamefont {Zhang}},\ and\ \bibinfo
  {author} {\bibfnamefont {S.}~\bibnamefont {Cao}},\ }\href
  {https://doi.org/10.1016/j.ceramint.2018.09.290} {\bibfield  {journal}
  {\bibinfo  {journal} {Ceramics International}\ }\textbf {\bibinfo {volume}
  {45}},\ \bibinfo {pages} {1093} (\bibinfo {year} {2019})}\BibitemShut
  {NoStop}%
\bibitem [{\citenamefont {Guo}\ \emph {et~al.}(2024)\citenamefont {Guo},
  \citenamefont {Campbell}, \citenamefont {Grutter}, \citenamefont {Noh},
  \citenamefont {Nan}, \citenamefont {Quarterman}, \citenamefont {Choi},
  \citenamefont {Tybell}, \citenamefont {Rzchowski},\ and\ \citenamefont
  {Eom}}]{PhysRevMaterials.8.L011401}%
  \BibitemOpen
  \bibfield  {author} {\bibinfo {author} {\bibfnamefont {L.}~\bibnamefont
  {Guo}}, \bibinfo {author} {\bibfnamefont {N.}~\bibnamefont {Campbell}},
  \bibinfo {author} {\bibfnamefont {A.~J.}\ \bibnamefont {Grutter}}, \bibinfo
  {author} {\bibfnamefont {G.}~\bibnamefont {Noh}}, \bibinfo {author}
  {\bibfnamefont {T.}~\bibnamefont {Nan}}, \bibinfo {author} {\bibfnamefont
  {P.}~\bibnamefont {Quarterman}}, \bibinfo {author} {\bibfnamefont {S.-Y.}\
  \bibnamefont {Choi}}, \bibinfo {author} {\bibfnamefont {T.}~\bibnamefont
  {Tybell}}, \bibinfo {author} {\bibfnamefont {M.~S.}\ \bibnamefont
  {Rzchowski}},\ and\ \bibinfo {author} {\bibfnamefont {C.-B.}\ \bibnamefont
  {Eom}},\ }\href {https://doi.org/10.1103/PhysRevMaterials.8.L011401}
  {\bibfield  {journal} {\bibinfo  {journal} {Phys. Rev. Mater.}\ }\textbf
  {\bibinfo {volume} {8}},\ \bibinfo {pages} {L011401} (\bibinfo {year}
  {2024})}\BibitemShut {NoStop}%
\bibitem [{\citenamefont {Balk}\ \emph {et~al.}(2017)\citenamefont {Balk},
  \citenamefont {Kim}, \citenamefont {Pierce}, \citenamefont {Stiles},
  \citenamefont {Unguris},\ and\ \citenamefont
  {Stavis}}]{PhysRevLett.119.077205}%
  \BibitemOpen
  \bibfield  {author} {\bibinfo {author} {\bibfnamefont {A.~L.}\ \bibnamefont
  {Balk}}, \bibinfo {author} {\bibfnamefont {K.-W.}\ \bibnamefont {Kim}},
  \bibinfo {author} {\bibfnamefont {D.~T.}\ \bibnamefont {Pierce}}, \bibinfo
  {author} {\bibfnamefont {M.~D.}\ \bibnamefont {Stiles}}, \bibinfo {author}
  {\bibfnamefont {J.}~\bibnamefont {Unguris}},\ and\ \bibinfo {author}
  {\bibfnamefont {S.~M.}\ \bibnamefont {Stavis}},\ }\href
  {https://doi.org/10.1103/PhysRevLett.119.077205} {\bibfield  {journal}
  {\bibinfo  {journal} {Phys. Rev. Lett.}\ }\textbf {\bibinfo {volume} {119}},\
  \bibinfo {pages} {077205} (\bibinfo {year} {2017})}\BibitemShut {NoStop}%
\bibitem [{\citenamefont {Belabbes}\ \emph {et~al.}(2016)\citenamefont
  {Belabbes}, \citenamefont {Bihlmayer}, \citenamefont {Bl{\"u}gel},\ and\
  \citenamefont {Manchon}}]{belabbes2016oxygen}%
  \BibitemOpen
  \bibfield  {author} {\bibinfo {author} {\bibfnamefont {A.}~\bibnamefont
  {Belabbes}}, \bibinfo {author} {\bibfnamefont {G.}~\bibnamefont {Bihlmayer}},
  \bibinfo {author} {\bibfnamefont {S.}~\bibnamefont {Bl{\"u}gel}},\ and\
  \bibinfo {author} {\bibfnamefont {A.}~\bibnamefont {Manchon}},\ }\href
  {https://doi.org/https://doi.org/10.1038/srep24634} {\bibfield  {journal}
  {\bibinfo  {journal} {Scientific reports}\ }\textbf {\bibinfo {volume} {6}},\
  \bibinfo {pages} {24634} (\bibinfo {year} {2016})}\BibitemShut {NoStop}%
\bibitem [{\citenamefont {Bhowmick}\ and\ \citenamefont
  {Sengupta}(2020)}]{dofhall}%
  \BibitemOpen
  \bibfield  {author} {\bibinfo {author} {\bibfnamefont {D.}~\bibnamefont
  {Bhowmick}}\ and\ \bibinfo {author} {\bibfnamefont {P.}~\bibnamefont
  {Sengupta}},\ }\href {https://doi.org/10.1103/PhysRevB.101.214403} {\bibfield
   {journal} {\bibinfo  {journal} {Phys. Rev. B}\ }\textbf {\bibinfo {volume}
  {101}},\ \bibinfo {pages} {214403} (\bibinfo {year} {2020})}\BibitemShut
  {NoStop}%
\bibitem [{\citenamefont {Chen}\ \emph {et~al.}(2022)\citenamefont {Chen},
  \citenamefont {Zhang}, \citenamefont {Liu}, \citenamefont {Hu}, \citenamefont
  {Shen},\ and\ \citenamefont {Sun}}]{PhysRevB.105.214428}%
  \BibitemOpen
  \bibfield  {author} {\bibinfo {author} {\bibfnamefont {X.}~\bibnamefont
  {Chen}}, \bibinfo {author} {\bibfnamefont {J.}~\bibnamefont {Zhang}},
  \bibinfo {author} {\bibfnamefont {B.}~\bibnamefont {Liu}}, \bibinfo {author}
  {\bibfnamefont {F.}~\bibnamefont {Hu}}, \bibinfo {author} {\bibfnamefont
  {B.}~\bibnamefont {Shen}},\ and\ \bibinfo {author} {\bibfnamefont
  {J.}~\bibnamefont {Sun}},\ }\href
  {https://doi.org/10.1103/PhysRevB.105.214428} {\bibfield  {journal} {\bibinfo
   {journal} {Phys. Rev. B}\ }\textbf {\bibinfo {volume} {105}},\ \bibinfo
  {pages} {214428} (\bibinfo {year} {2022})}\BibitemShut {NoStop}%
\bibitem [{We shall refer to the blue band as the flat band; however()}]{note}%
  \BibitemOpen
  We shall refer to the blue band as the flat band; however,\ \href@noop {}
  {\bibinfo {title} {if it becomes sufficiently dispersive due to interactions,
  it will be referred to as the middle band.}}\BibitemShut {Stop}%
\bibitem [{\citenamefont {Thouless}(1998)}]{thouless1998topological}%
  \BibitemOpen
  \bibfield  {author} {\bibinfo {author} {\bibfnamefont {D.}~\bibnamefont
  {Thouless}},\ }\href {https://doi.org/10.1142/3318} {\emph {\bibinfo {title}
  {Topological quantum numbers in nonrelativistic physics}}}\ (\bibinfo
  {publisher} {World Scientific},\ \bibinfo {year} {1998})\BibitemShut
  {NoStop}%
\bibitem [{\citenamefont {Mook}\ \emph
  {et~al.}(2014{\natexlab{b}})\citenamefont {Mook}, \citenamefont {Henk},\ and\
  \citenamefont {Mertig}}]{mook2014edge}%
  \BibitemOpen
  \bibfield  {author} {\bibinfo {author} {\bibfnamefont {A.}~\bibnamefont
  {Mook}}, \bibinfo {author} {\bibfnamefont {J.}~\bibnamefont {Henk}},\ and\
  \bibinfo {author} {\bibfnamefont {I.}~\bibnamefont {Mertig}},\ }\href
  {https://doi.org/10.1103/PhysRevB.90.024412} {\bibfield  {journal} {\bibinfo
  {journal} {Phys. Rev. B}\ }\textbf {\bibinfo {volume} {90}},\ \bibinfo
  {pages} {024412} (\bibinfo {year} {2014}{\natexlab{b}})}\BibitemShut
  {NoStop}%
\bibitem [{\citenamefont {Matsumoto}\ and\ \citenamefont
  {Murakami}(2011)}]{thermalhall}%
  \BibitemOpen
  \bibfield  {author} {\bibinfo {author} {\bibfnamefont {R.}~\bibnamefont
  {Matsumoto}}\ and\ \bibinfo {author} {\bibfnamefont {S.}~\bibnamefont
  {Murakami}},\ }\href {https://doi.org/10.1103/PhysRevB.84.184406} {\bibfield
  {journal} {\bibinfo  {journal} {Phys. Rev. B}\ }\textbf {\bibinfo {volume}
  {84}},\ \bibinfo {pages} {184406} (\bibinfo {year} {2011})}\BibitemShut
  {NoStop}%
\bibitem [{\citenamefont {Wojtkowiak}(1991)}]{Li2}%
  \BibitemOpen
  \bibfield  {author} {\bibinfo {author} {\bibfnamefont {Z.}~\bibnamefont
  {Wojtkowiak}},\ }\href {https://doi.org/http://dx.doi.org/10.1090/surv/037}
  {\bibfield  {journal} {\bibinfo  {journal} {Structural Properties of
  Polylogarithms, edited by Leonard Lewin, Amer. Math. Soc. Mathematical
  Surveys and Monographs}\ }\textbf {\bibinfo {volume} {37}},\ \bibinfo {pages}
  {205} (\bibinfo {year} {1991})}\BibitemShut {NoStop}%
\bibitem [{\citenamefont {Cao}\ \emph {et~al.}(2015)\citenamefont {Cao},
  \citenamefont {Chen},\ and\ \citenamefont {He}}]{cao2015magnon}%
  \BibitemOpen
  \bibfield  {author} {\bibinfo {author} {\bibfnamefont {X.}~\bibnamefont
  {Cao}}, \bibinfo {author} {\bibfnamefont {K.}~\bibnamefont {Chen}},\ and\
  \bibinfo {author} {\bibfnamefont {D.}~\bibnamefont {He}},\ }\href
  {https://doi.org/10.1088/0953-8984/27/16/166003} {\bibfield  {journal}
  {\bibinfo  {journal} {J. Phys.: Condens. Matter}\ }\textbf {\bibinfo {volume}
  {27}},\ \bibinfo {pages} {166003} (\bibinfo {year} {2015})}\BibitemShut
  {NoStop}%
\bibitem [{\citenamefont {Ideue}\ \emph {et~al.}(2012)\citenamefont {Ideue},
  \citenamefont {Onose}, \citenamefont {Katsura}, \citenamefont {Shiomi},
  \citenamefont {Ishiwata}, \citenamefont {Nagaosa},\ and\ \citenamefont
  {Tokura}}]{PhysRevB.85.134411}%
  \BibitemOpen
  \bibfield  {author} {\bibinfo {author} {\bibfnamefont {T.}~\bibnamefont
  {Ideue}}, \bibinfo {author} {\bibfnamefont {Y.}~\bibnamefont {Onose}},
  \bibinfo {author} {\bibfnamefont {H.}~\bibnamefont {Katsura}}, \bibinfo
  {author} {\bibfnamefont {Y.}~\bibnamefont {Shiomi}}, \bibinfo {author}
  {\bibfnamefont {S.}~\bibnamefont {Ishiwata}}, \bibinfo {author}
  {\bibfnamefont {N.}~\bibnamefont {Nagaosa}},\ and\ \bibinfo {author}
  {\bibfnamefont {Y.}~\bibnamefont {Tokura}},\ }\href
  {https://doi.org/10.1103/PhysRevB.85.134411} {\bibfield  {journal} {\bibinfo
  {journal} {Phys. Rev. B}\ }\textbf {\bibinfo {volume} {85}},\ \bibinfo
  {pages} {134411} (\bibinfo {year} {2012})}\BibitemShut {NoStop}%
\bibitem [{\citenamefont {Kitaev}(2006)}]{kitaev2006anyons}%
  \BibitemOpen
  \bibfield  {author} {\bibinfo {author} {\bibfnamefont {A.}~\bibnamefont
  {Kitaev}},\ }\href
  {https://doi.org/https://doi.org/10.1016/j.aop.2005.10.005} {\bibfield
  {journal} {\bibinfo  {journal} {Annals of Physics}\ }\textbf {\bibinfo
  {volume} {321}},\ \bibinfo {pages} {2} (\bibinfo {year} {2006})}\BibitemShut
  {NoStop}%
\end{thebibliography}%

\end{document}